\documentstyle[aaspp4,tighten,flushrt]{article}

\newcommand{\beq}{\begin{equation}}
\newcommand{\eeq}{\end{equation}}
\newcommand{\beqa}{\begin{eqnarray}}
\newcommand{\eeqa}{\end{eqnarray}}

\newcommand{\lexp}{\mathop{\langle}}
\newcommand{\rexp}{\mathop{\rangle}}

\def\d{\delta}
\def\dt{\tilde \delta}
\def\dD{\delta_{\rm D}}

\def\knl{k_{n\ell}}

\font\BF=cmmib10
\font\BFs=cmmib10 scaled 833
\def\k{{\hbox{\BF k}}}
\def\x{{\hbox{\BF x}}}
\def\ks{{\hbox{\BFs k}}}
\def\xs{{\hbox{\BFs x}}}
\def\q{{\hbox{\BF q}}}
\def\v{{\hbox{\BF v}}}

\def\la{\mathrel{\mathpalette\fun <}}
\def\ga{\mathrel{\mathpalette\fun >}}
\def\fun#1#2{\lower3.6pt\vbox{\baselineskip0pt\lineskip.9pt
        \ialign{$\mathsurround=0pt#1\hfill##\hfil$\crcr#2\crcr\sim\crcr}}}

\begin{document}

\lefthead{Scoccimarro et al.}
\righthead{Loop Corrections to the Bispectrum}
\hfill{\small CITA-97-10, FERMILAB-Pub-97/055-A}
\normalsize

\title{Non-linear Evolution of the Bispectrum of Cosmological Perturbations}

\author{Rom\'{a}n Scoccimarro$^{1,2,3}$, St\'ephane Colombi$^{3,4}$,
J. N. Fry$^{5}$, Joshua A. Frieman$^{2,6}$, \\
Eric Hivon$^{4,7}$, and Adrian Melott$^{8}$}

\vskip 1pc

\affil{${}^1$Department of Physics and Enrico Fermi Institute, 
University of Chicago, Chicago, IL 60637}

\affil{${}^2$NASA/Fermilab Astrophysics Center, Fermi National 
Accelerator Laboratory, \\ P.O. Box 500, Batavia, IL  60510}

\affil{${}^3$CITA, McLennan Physical Labs, 60 St George Street,
Toronto ON M5S 3H8, Canada}

\affil{${}^4$Institut d'Astrophysique de Paris, CNRS, 98 bis 
boulevard Arago, F-75014 Paris, France}

\affil{${}^5$Department of Physics, 215 Williamson Hall, 
University of Florida, Gainesville, FL  32611-8440}

\affil{${}^6$Department of Astronomy and Astrophysics, 
University of Chicago, Chicago, IL 60637}

\affil{${}^7$Theoretical Astrophysics Center, Juliane Maries Vej 30, 
DK-2100 Copenhagen, Denmark}

\affil{${}^8$Department of Physics and Astronomy, University of Kansas,
Lawrence, KS 66045}


%
\begin{abstract}
%
The bispectrum  $B(k_1, k_2, k_3)$, the three-point function of 
density fluctuations in Fourier space, is the lowest order statistic 
that carries information about the spatial coherence of large-scale 
structures.
For Gaussian initial conditions, when the density fluctuation amplitude 
is small ($\delta \ll 1$), tree-level (leading order) perturbation theory 
predicts a characteristic dependence of the bispectrum
on the shape of the triangle formed by the three wave vectors.
This configuration dependence provides a signature of gravitational 
instability, and departures from it in galaxy catalogs
can be interpreted as due to bias, that is, non-gravitational effects.
On the other hand, $N$-body simulations indicate that the reduced 
three-point function becomes relatively shape-independent in the 
strongly non-linear regime ($\delta \gg 1$).

In order to understand this non-linear transition and assess the domain 
of reliability of shape-dependence as a probe of bias, we calculate 
the one-loop (next-to-leading order) corrections to the bispectrum 
in perturbation theory.
We compare these results with measurements in numerical simulations 
with scale-free and Cold Dark Matter initial power spectra.
We find that the one-loop corrections account very well for the 
departures from the tree-level results measured in numerical
simulations on weakly non-linear scales ($\delta \la 1$).
In this regime, the reduced bispectrum qualitatively retains its
tree-level shape, but the amplitude can change significantly.
At smaller scales ($\delta \ga 1$), the reduced bispectrum in the 
simulations starts to flatten, an effect which 
can be partially understood from the one-loop results.
In the strong clustering regime, where perturbation theory
breaks down entirely, the simulation results confirm that the reduced
bispectrum has almost no dependence on triangle shape, in rough
agreement with the hierarchical ansatz.

\end{abstract}

\keywords{large-scale structure of universe; methods: numerical;
methods: statistical}

\clearpage 
%
%
\section{Introduction}
%
%

The growth of cosmological density fluctuations in perturbation theory 
(PT) is becoming a mature, well-understood subject, with techniques
established that in principle allow calculations to arbitrary order
(e.g., Goroff et al. 1986; Jain \& Bertschinger 1994).
On large scales, where the {\it rms} density fluctuations are small,
tree-level (leading order) PT gives the first non-vanishing contribution 
to statistical averages, and it has been used to understand the generation 
of higher order correlations in gravitational instability 
(e.g., Peebles 1980; Fry 1984; Juszkiewicz, Bouchet \& Colombi 1993; 
Bernardeau 1992, 1994).
Comparison with fully non-linear numerical simulations has shown
this perturbative approach to be very successful 
(Juszkiewicz et al. 1993, 1995; Lucchin et al. 1994; 
Bernardeau 1994; Fry 1994a; {\L}okas et al.~1995; 
Gazta\~{n}aga \& Baugh 1995; Baugh, Gazta\~{n}aga \& Efstathiou 1995).

On smaller scales, where the density fluctuation amplitude approaches 
or exceeds unity, loop (next-to-leading and higher order)
corrections to the tree-level PT results should become important.
The question then arises of whether our understanding of
clustering can be extended from large scales further into the
non-linear regime by using one-loop PT.
In previous papers, we considered one-loop corrections to one-point 
cumulants of unsmoothed fields (Scoccimarro \& Frieman 1996), 
the power spectrum, variance, and two-point correlation function 
(Scoccimarro \& Frieman 1996b; see also Makino, Sasaki, \& Suto 1992; 
{\L}okas et al.~1996), and the bispectrum and skewness including
smoothing effects (Scoccimarro 1997).

In this paper, we present one-loop perturbative bispectra for a 
variety of initial power spectra, and we compare them with results of 
numerical simulations.
The bispectrum $B(k_1, k_2, k_3)$, the Fourier transform of the 
connected 
three-point correlation function of the density field perturbations, 
is of interest for several reasons.
For Gaussian initial conditions, the connected $N-$point correlation 
functions vanish in linear theory for $N>2$.
The bispectrum is therefore intrinsically
non-linear, and it is the lowest-order statistic with this property.
It is also the lowest-order statistic that carries information about
the spatial coherence of the density and velocity fields; by contrast,
the density power spectrum $ P(k) \sim \lexp |\delta(\k)|^2 \rexp $ 
is independent of phase correlations.

In addition,
the dependence of the tree-level bispectrum on the shape of the triangle
formed by the three wave vectors $\k_1, \k_2, \k_3$ is a characteristic
signature of gravitational instability.
The degree to which the observed {\it galaxy} distribution exhibits 
this predicted configuration dependence provides an independent 
probe of bias, that is, of the relation
between the galaxy and mass density distributions (Fry 1994b).
In the galaxy data that have been studied to date, the
Shane-Wirtanen (Lick) angular catalog, the reduced bispectrum was not
found to exhibit the expected tree-level dependence on configuration shape
(Fry \& Seldner 1982), and this can be interpreted as a possible
indication that these galaxies are biased relative to the mass (Fry 1994b).
However, numerical simulations indicate that the reduced bispectrum 
becomes much less dependent on configuration 
(Fry, Melott, \& Shandarin 1993, hereafter FMS; Fry, Melott, \&
Shandarin 1995)  in the highly non-linear regime.
Thus, in order to rigorously apply this test, i.e., to use the bispectrum 
shape as a quantitative probe of bias, we must assess the reliability of 
the tree-level predictions and quantify where and how they break down.
One-loop PT, in conjunction with $N$-body simulations, provides
a framework for accomplishing this.

Study of the one-loop bispectrum is also of practical value 
for the analysis of galaxy surveys.
For scale-free initial power spectra, $P(k) \propto k^n $,
measurement uncertainties in the reduced bispectrum scale as 
$ (\knl/k)^{(n+3)/2} $ (FMS), where $k$ is an inverse wavelength and 
$ \knl $ is the scale of nonlinearity (see eq.~[\ref{knldef}] below).
Therefore, measuring the reduced bispectrum is difficult on 
large scales ($k \ll \knl$).
For better accuracy, we would like to measure $B$ on scales with $k$ 
as large as possible, at least not much smaller than $ \knl $.
Since the tree-level PT result is not expected to be accurate in the 
strong clustering regime, $k \ga \knl$, we would like to know 
the size and nature of the corrections to the tree-level prediction 
on scales $k$ comparable to $ \knl $. This is precisely the domain
where one-loop PT can be very helpful, and can be additionally checked
against numerical simulations which are reliable at these scales. 

The paper is organized as follows.
In Section~2, we briefly review PT solutions and the loop expansion of
the power spectrum and bispectrum. In Section~3, we apply PT to power-law
initial power spectra and give some analytic results.
Section~4 describes the numerical simulations analyzed in this work and 
Section~5 compares them to the one-loop perturbative results on the
power spectrum and bispectrum.
Section~6 contains a final discussion.
Finally, Appendix A describes the method used to measure the
bispectrum in numerical simulations, and Appendix B reviews the
general framework of PT, including the extension to
arbitrary cosmological parameters $\Omega$ and $\Lambda$.

%
%
\section{Perturbation Theory}
%
%
We work in transform space with the Fourier amplitude of the
density contrast $ \d(\x) = [\rho(\x,t) - \bar \rho] / \bar \rho $, defined
such that
\beq
\dt(\k) = \int {d^3 x \over (2 \pi)^3} \, \d(\x) \, e^{-i\ks\cdot\xs} .
\eeq
To linear order, $ \dt(\k,t) = a(t) \, \delta_1(\k)  $, where 
$a(t)$ is the cosmic scale factor (normalized to $a(t_0)=1$ today), 
and $\delta_1(\k)$ denotes the linear density fluctuation amplitude 
(at the present epoch with the scale factor normalization above).
We consider an Einstein-de Sitter Universe, with density parameter
$\Omega=1$, in which case the PT expansion for the density contrast
can be written
\beq
	\tilde{\delta}(\k,t) = \sum_{n=1}^{\infty} a^n(t) \, 
	\delta_n(\k)  \label{ptansatz}.
\eeq
Modeling the matter as pressureless non-relativistic `dust',
an appropriate description for cold dark matter before shell crossing,
the fluid equations of motion (see Appendix~\ref{eom}) 
determine $\delta_n(\k) $
in terms of the linear fluctuations,
\beq
	\delta_n(\k) = \int d^3q_1 \ldots \int d^3q_n \; 
\dD(\k - \q_1 - \cdots - \q_n) \, F_n^{(s)} (\q_1, \ldots, \q_n)
	\, \delta_1(\q_1) \cdots \delta_1(\q_n)  \label{ec:deltan},
\eeq
where the $F_n^{(s)}$ are dimensionless, symmetric, scalar 
functions of the wave vectors $ \{ \q_1,  \ldots , \q_n \} $
(Goroff et al. 1986; Jain \& Bertschinger 1994) 
and $ \dD $ is the Dirac $\d$-function.
The recursion relations from which these kernels can be derived 
are provided for reference in Appendix B.

A systematic framework for  calculating  correlations of cosmological
fields in PT has been formulated using
diagrammatic techniques (Fry 1984; Goroff et al.~1986; Wise 1988;
Scoccimarro \& Frieman 1996). From this point of view, leading order
PT for the statistical quantities of interest
corresponds to tree graphs, next-to-leading order PT contributions
can be described in terms of one-loop graphs, etc.

The simplest statistic of interest is the power spectrum $P(k)$, 
the second moment of the Fourier amplitude of the density contrast, 
defined by
\beq
\lexp \dt(\k) \dt(\k') \rexp = \dD (\k+\k') \,  P(k) .
\eeq
By statistical isotropy, the power spectrum depends only
on the magnitude of $\k$.
To tree level (linear theory), the power spectrum keeps its shape 
and simply grows by an overall amplitude.
One-loop (first non-linear) corrections introduce coupling between 
different Fourier modes and increase or decrease the growth rates 
relative to linear theory,
depending on the amount of small-scale power in the initial spectrum
(Klypin \& Melott 1992; Makino et al.~1992; \L okas et al. 1996;
Scoccimarro \& Frieman 1996b).
We can write the loop expansion for $P(k)$ up to one-loop corrections as
(henceforth we suppress the implicit time dependence)
\beq
	P(k) = P^{(0)}(k) + P^{(1)}(k) + \cdots \, .
	\label{Ploopexp}
\eeq
The superscript $(n)$ denotes an $n$-loop contribution.
The tree-level ($0$-loop) contribution is just the linear spectrum, with
\beq
a^2 \lexp \delta_1(\k) \delta_1(\k') \rexp =
 \dD(\k+\k') \,  P^{(0)}(k)  \label{P^(0)}, 
\eeq
and the one-loop contribution consists of two terms,
\beq
	P^{(1)}(k) = P_{22}(k) + P_{13}(k)
	\label{P^(1)},
\eeq
where 
\begin{eqnarray}
	P_{22}(k) & = &  2 \int [ F_2^{(s)}(\k-\q,\q) ]^2
	\, P^{(0)}(|\k-\q|) \, P^{(0)}(q) \, d^3q \label{P22},  \\
	P_{13}(k)  & = &   6  \int F_3^{(s)}(\k,\q,-\q) 
	\, P^{(0)}(k) \, P^{(0)}(q) \, d^3q  \label{P13}.
\end{eqnarray}
Here $P_{ij}$ denotes the
amplitude corresponding to the contribution
$\lexp \delta_i(\k)  \delta_j(\k) \rangle$ to the power spectrum.
We have assumed Gaussian initial conditions, for which
$P_{ij}$ vanishes if $i+j$ is odd.

The third moment in the Fourier domain gives the bispectrum, 
$ B(k_1, k_2, k_3) $, defined by
\beq
\lexp \dt(\k_1) \dt(\k_2) \dt(\k_3) \rexp
= \dD(\k_1+\k_2+\k_3) \, B(k_1,k_2,k_3) .
\eeq
The Dirac $\delta$-function ensures that the bispectrum is defined 
only for configurations that form closed triangles, 
$ \sum \k_i = 0 $.
For Gaussian initial conditions, the first non-vanishing contribution 
to the connected $n$-point correlation function requires PT to order 
$n-1$ (Fry 1984).
The loop expansion for the bispectrum reads (Scoccimarro 1997):
\beq
B(k_1,k_2,k_3) = B^{(0)}(k_1,k_2,k_3) +
B^{(1)}(k_1,k_2,k_3) + \cdots .  \label{Bloopexp}
\eeq
The tree-level term is
\begin{eqnarray}
B^{(0)}(k_1,k_2,k_3) &\equiv&
2 \, P^{(0)}(k_1) \, P^{(0)}(k_2) \, F_2^{(s)}(\k_1,\k_2) +
2 \, P^{(0)}(k_2) \, P^{(0)}(k_3) \, F_2^{(s)}(\k_2,\k_3)
\nonumber \\	 & &  +
2 \, P^{(0)}(k_3) \, P^{(0)}(k_1) \, F_2^{(s)}(\k_3,\k_1),
	\label{Btree}
\end{eqnarray}
where the second-order kernel for the Einstein-de Sitter model is
\beq
2F_2^{(s)}(\k_i,\k_j) = {10 \over 7} + \hat \k_i \cdot \hat \k_j
\left( {k_i \over k_j}+ {k_j \over k_i} \right) + {4 \over 7}
(\hat \k_i \cdot \hat \k_j)^2 \, ; \label{Qij}
\eeq
hats denote unit vectors (Fry 1984).
For $ \Omega \ne 1 $, and vanishing cosmological constant, 
the factors $ 10/7 $ and $ 4/7 $ become
$ 1 + \kappa $ and $ 1 - \kappa $, where
$ \kappa \approx \frac{3}{7} \Omega^{-2/63} $ depends only very
weakly on $ \Omega $ (Bouchet et al.~1995, Hivon et al.~1995). A
discussion of the dependence of PT kernels on cosmological parameters
is presented in Appendix B.3. 

The one-loop bispectrum comprises several terms,
\beq
B^{(1)}(k_1,k_2,k_3) = B_{222}(k_1,k_2,k_3) 
 + B_{321}^{I}(k_1,k_2,k_3) + B_{321}^{II}(k_1,k_2,k_3) 
 + B_{411}(k_1,k_2,k_3)  \label{B1loop},
\eeq
with:

\label{1loopB}
\begin{eqnarray}
B_{222} & = & 8 \int d^{3}q \, P^{(0)}(q) \, 
	 F_2^{(s)}(-\q,\q+\k_1) \, P^{(0)}(|\q+\k_1|)
	 \, F_2^{(s)}(-\q- \k_1,\q-\k_2) \nonumber \\
	 && \qquad \times P^{(0)}(|\q-\k_2|) 
	\, F_2^{(s)}(\k_2-\q,\q) , \label{B222} \\
B_{321}^I & = & 6 \, P^{(0)}(k_3) \int d^{3}q \, P^{(0)}(q)
	 \, F_3^{(s)}(-\q,\q-\k_2,-\k_3 )
	 \, P^{(0)}(|\q-\k_2|) \nonumber \\
	 &&   \qquad \times F_2^{(s)}(\q,\k_2-\q) 
	+ \, {\rm 5 \ permutations} ,  \label{B321I} \\
B_{321}^{II} & = & 6
	 \, P^{(0)}(k_2) \, P^{(0)}(k_3) \, F_2^{(s)}(\k_2,\k_3)
	\int d^{3}q \, P^{(0)}(q) 
    \, F_3^{(s)}(\k_3,\q,-\q) \nonumber \\
	 &&  \qquad + \, {\rm 5 \ permutations}
	\label{B321II}, \\
B_{411} & = & 12  \, P^{(0)}(k_2)  \, P^{(0)}(k_3)
	\int d^{3}q \, P^{(0)}(q)
    \, F_4^{(s)}(\q,-\q,-\k_2,-\k_3) \nonumber \\
	 &&  \qquad  + \, {\rm 2 \ permutations} \label{B411}.
\end{eqnarray}
Again, the subscript $ijk$ denotes a contribution of order
$\lexp \delta_i \delta_j \delta_k \rexp$ to the bispectrum. The reader
is referred to  Scoccimarro (1997), Fig.~3, for a diagrammatic representation
of these terms.

The reduced bispectrum, or hierarchical three-point amplitude $Q$ 
is defined as
\beq
Q(k_1,k_2,k_3) = { B(k_1,k_2,k_3) \over 
P(k_1) \, P(k_2) + P(k_2) \, P(k_3) + P(k_3) \, P(k_1) } . \label{q}
\eeq
The loop expansion of the numerator and denominator yields:
\beq
 Q = \frac{ B^{(0)}(k_1,k_2,k_3) + B^{(1)}(k_1,k_2,k_3) + \cdots}
 {\Sigma^{(0)}(k_1,k_2,k_3) + \Sigma^{(1)}(k_1,k_2,k_3) + \cdots} 
  \,  ,   \label{Qratio}
\eeq
where 
\beqa
\Sigma^{(0)}(k_1,k_2,k_3) &=&  P^{(0)}(k_1) P^{(0)}(k_2) +
     P^{(0)}(k_2) P^{(0)}(k_3) +  P^{(0)}(k_3) P^{(0)}(k_1)
	\label{sigma0},  \\
\Sigma^{(1)}(k_1,k_2,k_3) &=&  P^{(0)}(k_1) P^{(1)}(k_2) +
	 {\rm 5 \ permutations}  \label{sigma1}.
\eeqa
The loop expansion of
$Q \equiv  Q^{(0)} + Q^{(1)} +\cdots $ gives 
\beqa
Q^{(0)} &=& \frac{ B^{(0)}(k_1,k_2,k_3)}
    { \Sigma^{(0)}(k_1,k_2,k_3) } \label{qtree}, \\
Q^{(1)}  &=&  \frac{B^{(1)}}{\Sigma^{(0)}}
    - \frac{ Q^{(0)} \Sigma^{(1)}}{\Sigma^{(0)}} \label{q1l}.
\eeqa
Note that $Q^{(1)}$ depends on the normalization of the 
linear power spectrum, and its amplitude increases with time evolution.
On the other hand, from equations (\ref{Btree}), (\ref{sigma0}), and 
(\ref{qtree}) it follows that $Q^{(0)}$ is independent of time and 
normalization (Fry 1984).
Furthermore, for scale-free initial conditions, $P^{(0)}(k) \propto k^{n}$, 
$Q^{(0)}$ is also independent of overall scale.
For the particular case of equilateral configurations 
($ k_{1}=k_{2}=k_{3} $ and $ \hat{\k_{i}} \cdot \hat{\k_{j}}=-0.5 $ 
for all pairs), $Q^{(0)}$ is independent of spectral index as 
well, $Q^{(0)}_{\rm EQ}=4/7$.
In general, for scale-free initial power spectra,
$Q^{(0)}$ depends on configuration shape through, e.g., the ratio
$k_{1}/k_{2}$ and the angle  $\theta$  defined by $\hat{\k}_{1} \cdot
\hat{\k}_{2} =\cos{\theta}$.
This configuration dependence of $Q^{(0)}$ reflects the anisotropy of
structures and flows generated by gravitational instability.
From equations (\ref{qtree}), (\ref{sigma0}), and (\ref{Btree}), 
it follows that $Q^{(0)}$ would be independent of configuration shape 
if $F_2^{(s)}(\k_i,\k_j)$ (see eq.~[\ref{Qij}]) were a constant.
The configuration dependence of $F_2^{(s)}(\k_i,\k_j)$ implied 
by equation (\ref{Qij}) has two sources: the term linear in
$\hat{\k}_{i} \cdot \hat{\k}_{j}$ comes from the
gradients of the density field in the direction of the flow, whereas
the term quadratic in $\hat{\k}_{i} \cdot \hat{\k}_{j}$ represents
the gradients of the velocity field in the direction of the flow
(Scoccimarro 1997).
Thus, $Q^{(0)}$ is enhanced if the wave vectors are collinear 
($\theta=0, \pi$), which  reflects the fact that large-scale flows 
generated by gravitational instability are  mostly parallel to density
gradients (see, e.g., dotted lines in Fig.~1). 
This physical interpretation provides some insight into what is 
expected to happen as the transition to the non-linear regime is made.
As long as the evolution of structures is dominated by large-scale 
motions, the shape dependence of $Q$ should remain qualitatively 
the same as at tree level. In fact, for spectra dominated by large
scale power, the anisotropy of structures is amplified by the 
``pancaking'' process described by the Zel'dovich (1970) approximation 
(Coles et al. 1993; Melott \& Shandarin 1993) and this would lead to
an enhancement of the configuration dependence of $Q$. On the other
hand, when substantial velocity dispersion   
develops on small scales due to virialization, the interpretation
above suggests that one 
should see a flattening in $Q$: non-collinear configurations become 
more probable, due to the loss of coherence of structures and flows 
(i.e., the gradient terms in  eq.~[\ref{Qij}] 
are smoothed out due to random motions.)

To characterize the degree of non-linear evolution when including
one-loop corrections to the power spectrum and bispectrum, it is
convenient to define a physical scale from the linear power spectrum.
One such scale is the correlation length $R_0$, the scale on which the 
smoothed linear variance is unity, $\sigma^2_\ell(R_0) \equiv 1$.
The variance is defined by 
\beq
\sigma^2_\ell(R)  = \int d^{3}k \, P^{(0)}(k) \, W^{2}(k R)
\label{R0},
\eeq
where $W(x)$ is the Fourier transform of a window function 
(usually a top-hat or Gaussian) of characteristic scale $R$.
For scale-free initial power spectra, 
$P^{(0)}(k) = A a^2 k^n$, the variance scales as $\sigma^2_\ell(R) = 
(R/R_0)^{-(n+3)}$. For Gaussian smoothing, $W(kR)=\exp(-k^2R^2/2)$,
the linear correlation length satisfies
\beq
R_{0}^{n+3} \equiv 2\pi \, A \, a^{2} \, 
\Gamma \left( \frac{n+3}{2} \right) . \label{R0pl}
\eeq
In the $128^3$ PM 
simulations described below the epoch of evolution is labeled by 
$ \knl $, the wave number that is on the threshold of going nonlinear 
as determined in linear theory, defined by 
\beq
 \int_{k_f}^{\knl} d^3 k \, P^{(0)}(k) 
= {4\pi \over (n+3)} \, A \, a^2 k^{n+3}_{n\ell} \equiv 1,  
\label{knldef}
\eeq
where $k_f\equiv 2\pi/L$ is the fundamental mode of the simulation
box of  side $L$. Equations (\ref{R0pl}) and
(\ref{knldef}) imply that $\knl R_0 \cong \Gamma[(n+5)/2]$.

%
%
\section{One-Loop Results for the Bispectrum}
%
%
For scale-free initial power spectra, the one-loop integrals 
in equations (\ref{B222})--(\ref{B411}) can be calculated
analytically in the range $-3<n<-1$ by using dimensional 
regularization (Scoccimarro 1997).
In this spectral range, the resulting bispectrum obeys self-similarity 
and, based on previous results for the power spectrum 
(Scoccimarro \& Frieman 1996b), the one-loop calculations are
expected to give a good description of the transition to the
nonlinear regime.

Due to statistical homogeneity and isotropy, the bispectrum 
$B (\k_{1},\k_{2},\k_{3})$ depends on time, the magnitudes 
$k_{1}$, $k_{2}$, and the angle $\theta$ 
($\hat{\k}_{1}\cdot \hat{\k}_{2} \equiv \cos{\theta}$).
In order to display the analytic results, however,
it is more convenient to trade the
variable $\theta$ for the third side of the triangle, 
$k_{3} = |\k_1 + \k_2 |$.
Let $B^{(1)}(\k_{1},\k_{2},\k_{3}) \equiv A^{3} a^{6}
\pi^{3} \, b^{(1)}(k_1,k_2,k_3)$, with $\k_1 + \k_2 +\k_3 \equiv 0$.
Then, for $n=-2$ it  follows (Scoccimarro 1997):
\begin{eqnarray}
	 b^{(1)}(k_1,k_2,k_3) & = & -{{30279}\over {34496\,{{k_1}^3}}} -
{{2635\,{{k_1}^2}}\over {51744\,{{k_2}^5}}} -
{{37313\,k_1}\over {206976\,{{k_2}^4}}} +
{{38431}\over {68992\,k_1\,{{k_2}^2}}}
	\nonumber   \\
	 &  & +
{{233\,{{k_1}^6}}\over {8624\,{{k_2}^4}\,{{k_3}^5}}} -
{{16517\,{{k_1}^5}}\over
    {362208\,{{k_2}^3}\,{{k_3}^5}}} +
{{197\,{{k_1}^4}}\over {7392\,{{k_2}^2}\,{{k_3}^5}}} -
{{78691\,{{k_1}^3}}\over {275968\,k_2\,{{k_3}^5}}}
	\nonumber  \\
	 &  & -
{{23\,{{k_1}^5}}\over
    {103488\,{{k_2}^4}\,{{k_3}^4}}} +
{{9791\,{{k_1}^4}}\over
    {206976\,{{k_2}^3}\,{{k_3}^4}}} +
{{703\,{{k_1}^3}}\over
    {68992\,{{k_2}^2}\,{{k_3}^4}}} +
{{19867\,{{k_1}^2}}\over {206976\,k_2\,{{k_3}^4}}}
	\nonumber  \\
	 &  &  +
{{5311\,{{k_1}^2}}\over
    {34496\,{{k_2}^2}\,{{k_3}^3}}} +
{{42983\,k_1}\over {362208\,k_2\,{{k_3}^3}}}     +
{{131\,k_1}\over {3696\,{{k_2}^2}\,{{k_3}^2}}} +
{{28393}\over {19712\,k_1\,k_2\,k_3}}
\nonumber  \\
	 &  &  +
{{53973\,{{k_1}^7}}\over
    {1931776\,{{k_2}^5}\,{{k_3}^5}}}+
{{108685\,k_1\,k_2}\over {181104\,{{k_3}^5}}}+
{{59599\,{{k_1}^3}}\over
    {362208\,{{k_2}^3}\,{{k_3}^3}}}
    \nonumber  \\
	 &  &  +\ {\rm permutations}.
	\label{bispecnm2}
\end{eqnarray}
A simple result can be obtained for equilateral configurations.
Given that the one-loop power spectrum for $n=-2$ can be written as
$P^{(1)}(k)=A^2 a^4 55 \pi^3 /(98 k)$ (Scoccimarro \& Frieman 1996b), the
hierarchical amplitude for equilateral configurations 
at the one-loop level is:
\beq
Q_{\rm EQ}(n=-2)= \frac{4}{7} +
\frac{1426697 }{3863552} \pi^{3/2} \ kR_0 = 0.57 + 2.06 \ kR_0.
\label{Qeqnm2}
\eeq
For $n=-1.5$, the corresponding result reads:
\beq
Q_{\rm EQ}(n=-1.5) = 0.57 + 1.32 \ (kR_0)^{3/2}
\label{Qeqnm1p5}.
\eeq

For other spectral indices in the range $-3<n<-1$, 
the one-loop bispectrum can be expressed in terms of 
hypergeometric functions (Scoccimarro~1997).
On the other hand, when $n \geq -1$, one-loop PT leads to ultraviolet 
($k \rightarrow \infty$) divergences that must be regulated by the 
introduction of a cutoff scale.
In this case we take the initial power spectrum to be
$P^{(0)}(k) = A \ a^{2} \ k^{n}$ for $\epsilon < k < k_{c}$ and zero
otherwise. For convenience in comparison with the $n=-1$, $0$, $1$
numerical simulations described below, in these cases we take
$ k_1 = 1 $, $ k_2 = 1/2 $, $ \epsilon = 1/16 $, and $ k_c = 4 $.
The integration  of equations (\ref{B222})--(\ref{B411}) is then done 
numerically.
Given the complexity of the calculations involved, it is desirable 
to verify that the numerical integration code is correct.
For this reason, we have written two completely independent
codes of numerical integration, one based on Romberg integration, 
the other using Gaussian adaptive integration.
The results of both codes agree with each other very well 
over the whole range of spectral indices and cutoff
parameters considered.
They also agree with the analytic result of equation (\ref{bispecnm2}) 
for the case $n=-2$, in the limit that the spectral cutoffs are removed.
The one-loop bispectrum for cold dark matter
(CDM) initial spectra does not present any additional complications for
the numerical evaluation and is calculated using the same
program. Since in this case $P^{(0)} \approx k^{-3}$ at large $k$,
there are no ultraviolet divergences for this spectrum.
Typically, the numerical evaluation of the one-loop bispectrum  requires 
a few hours in one processor of a Silicon Graphics Power Challenge or
DEC-Alpha 2100 workstation.

\clearpage 
%
%
\section{Numerical Simulations}
%
%
%
\subsection{Scale-Free Simulations}
%

We compare our perturbative calculations to numerical results from
two sets of scale-free simulations with Gaussian initial conditions.

The first is an ensemble of simulations with initial spectral
indices $n=-2$, $-1$, $0$ and $+1$, performed by Melott \& Shandarin
(1993) with the Particle-Mesh (PM) code of Melott (1986).
These simulations involve $N_{\rm par}=128^3$ particles and 
a $128^3$ staggered mesh.
The force on a cell is a result of differencing the potential 
in its 8 neighbors.
These PM simulations have about twice the usual resolution by using 
a staggered mesh scheme (Melott 1986; Melott, Weinberg \& Gott 1988).
Models with $ n = +1 $ and $-1$ have an initial amplitude such
that the rms fluctuation averaged over one mesh cell is
$ (\Delta \rho / \rho)_0 = 0.05 $.
Models with $ n = 0 $ and $-2$ have the slightly larger initial 
amplitude ($\Delta\rho/\rho)_0=0.25$.
The power spectrum and bispectrum data we use here were
measured in these simulations respectively by 
Melott \& Shandarin (1993) and by FMS.
Four independent realizations were generated for each value of $n$, 
using the same sets of random number seeds.
Besides improving the signal, averaging over four realizations 
allows us to estimate fairly the uncertainties in our results.
For these simulations, the epoch of evolution is labeled by
$ \knl $, the wave number that is on the threshold of going nonlinear
as determined in linear theory given in equation (\ref{knldef}).
The simulation output times are characterized by 
$ \knl = 64 $, 32, 16, 8, and 4, 
or $ k_{ny} / \knl = 1 $, 2, 4, 8, 16, 
where $k_{ny}=64$ is the Nyquist frequency of the simulation 
(wave numbers given in units of the fundamental mode of 
the simulation box of side $L$, $k_f\equiv 2\pi/L$).

We have also analyzed two simulations, one with $n=-2$ (already used
in \L okas et al.~1996) and the other
one with $n=-1.5$, done with a vectorized PM code
(Moutarde at al.~1991) modified to run in parallel on several
processors of a CRAY-98 (Hivon 1995). They involve $256^{3}$
particles and use a $256^{3}$ (unstaggered) mesh to compute the forces.
The initial conditions were set by using the
Zel'dovich approximation on a ``glass'' (see, e.g., White 1994), and the
initial power spectrum is given by
$P(k)=A^2(n) \ (k/k_{ny})^n/256^3$, where $k_{ny}$ is the Nyquist frequency
of the particles in units of the fundamental mode, and $A(-2)=0.2$, $A(-1.5)=
1/\sqrt{128}$. For $n=-2$, we have analyzed outputs with $a=8, 11.31,
16, 22.63$, whereas for $n=-1.5$ we have analyzed
outputs with $a=22.63, 32, 45.25, 64$, where both simulations start
at $a=1$.
To avoid spurious effects (Melott et al. 1988;  
Kauffman \& Melott 1992; Colombi, Bouchet \& Schaeffer 1994, 1995), we
only consider scales $k$ that  
satisfy the requirement $4 \leq k_1/k_f \leq k_{ny}/2$, where
$k_f \equiv 2\pi /L$ is the fundamental mode of the simulation box.

%
\subsection{CDM Simulations}
%

The CDM simulation we analyzed was done by Couchman, Thomas \&
Pearce (1995) with an adaptive P$^{3}$M
(Particle-Particle-Particle-Mesh) code and involves $128^{3}$  
particles in a box of length $ 100 \, h^{-1} \, {\rm  Mpc} $
($ h \equiv H_0/100 \, {\rm km \,  s^{-1} \, Mpc^{-1}} $, 
where $H_0$ is the Hubble constant).  
These simulation data are publicly available through the Hydra Consortium 
Web page ({\sf http://coho.astro.uwo.ca/pub/consort.html}).
They correspond to an $\Omega=1$ model, with linear CDM
power spectrum characterized by a shape parameter
$ \Gamma=\Omega h = 0.25 $, in approximate agreement
with the observed galaxy power spectrum on large scales
(e.g., Peacock \& Dodds 1994).
The initial conditions were set by using the
Zel'dovich approximation on a grid. We have analyzed output times
at which the linear rms density fluctuation amplitude
in top-hat spheres of radius $ R = 8 h^{-1} \, {\rm Mpc} $ is given by
$\sigma_{8}=0.2057$, 0.3291, 0.64. These times correspond respectively
to scale factors $a=0.3214$, 0.5143, 1, where $a=0.02$ initially.
For the measurements, we have only considered scales in the range
$4 \leq k/k_f \leq 50$. In particular, the $k_1/k_2=2$ configurations
shown in the figures below correspond to $k_1/k_f = 15$, 30, 40.

%
%
\section{One-Loop Perturbation Theory vs. Numerical Simulations}
\begin{figure}[t!]
\centering
\centerline{\epsfxsize=18truecm\epsfysize=18truecm\epsfbox{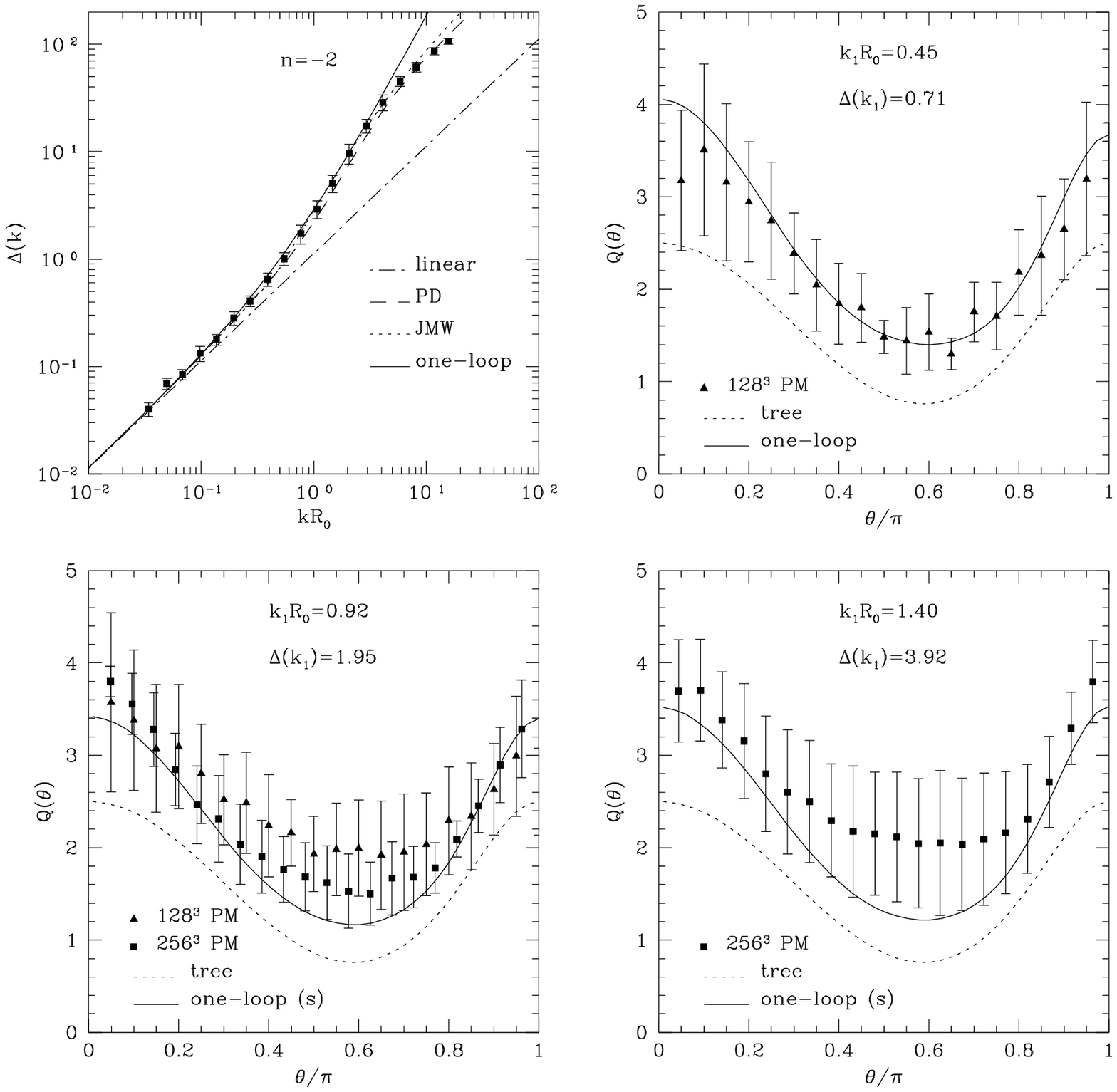}}
\caption{The left top panel shows the non-linear power spectrum in
terms of  $\Delta(k)\equiv 4\pi k^3 P(k)$ as a 
function of scale for $n=-2$ scale-free initial conditions. 
Symbols denote measurements in numerical simulations, whereas lines show 
the linear, PD, JMW and one-loop perturbative results, as indicated.
The other three panels show the hierarchical amplitude $Q$ 
for triangle configurations with $k_{1}/k_{2}=2$, as a function 
of the angle $\theta$ between $\hat{\k}_{1}$ and $\hat{\k}_{2}$, 
in numerical simulations and for tree-level and one-loop PT.
The panels correspond to stages of non-linear evolution characterized 
by $\Delta(k_1)= 0.71$, 1.95, 3.92.
}
\label{fig1}
\end{figure}

%
%
We now compare the one-loop perturbative predictions with the $N$-body
results for scale-free and CDM power spectra.
Figure~\ref{fig1} presents our main results for the $n=-2$ spectrum.
On the top left panel, we show the contribution per logarithmic
wave number interval to the variance,
$\Delta(k) \equiv 4 \pi k^3 P(k)$, as a function
of scale. In linear theory, from equation (\ref{knldef}), 
$ \Delta^{(0)}(k) = (n+3)(k/\knl)^{n+3} $.
The symbols correspond to the $256^3$ $N$-body results averaged over
the four different time outputs, assuming self-similar evolution.
The error bars in this plot are calculated from the dispersion in 
this averaging procedure (see Appendix A.2 for details).
In the same plot, we include the linear theory extrapolation, 
$\Delta^{(0)} (k;n=-2) = (k/\knl) $, the one-loop PT prediction, 
and the phenomenological $N$-body fitting formulae proposed by 
Jain, Mo \& White (1995, JMW) and by Peacock \& Dodds (1996, PD), 
based on earlier work by Hamilton et al.~(1991).
Note how well the one-loop analytic result,
\beq
\Delta(kR_0) = (2/\sqrt{\pi}) \ kR_0 \
\Big( 1 + \frac{55}{196}\pi^{3/2} \ kR_0
\Big) = 1.128 \ kR_0 \
\Big( 1 + 1.562 \ kR_0
\Big)
\label{Deltanm2},
\eeq
describes the numerical simulation measurements and fitting formulae
up to scales where $\Delta \approx 30$, where the power spectrum goes 
over to the stable clustering regime.
This is remarkable, given the simplicity of equation (\ref{Deltanm2}) 
when compared to the JMW and PD fitting formulae.

In the other three panels in Figure~\ref{fig1}, we
show results for the hierarchical amplitude $Q$ for triangle configurations
with $k_1/k_2=2$, as a function of the angle $\theta$ between $\k_1$ and
$\k_2$, for three different scales corresponding to $\Delta(k_1)=0.71,
1.95, 3.92$. Filled squares denote averages over the four
simulation output times as mentioned
above for the $256^3$ PM code. Filled triangles denote results taken from the
$128^3$ PM simulation. In this case, the displayed values correspond to the
average over the four different realizations, using only a single
output time for each one. The error bars correspond
to the dispersion over the measurements (see
Appendix A). They should be more reliable
than those estimated from the $256^3$ simulations, which may also
reflect departures from self-similarity.

We note the clear departure of the $N$-body results from the 
tree-level PT prediction equation (\ref{qtree}) at  $k_1R_0=0.45$ and 
the very good agreement when the one-loop correction 
(eq.~[\ref{q1l}]) is included.
The net effect of non-linear evolution at this stage is to partially 
increase the configuration dependence of $Q$, as expected from the fact 
that for this initial spectrum the evolution on weakly non-linear scales 
is still dominated by large-scale coherent motions.
When $\Delta(k_1) > 1$, it is not justified to expand the denominator 
in equation (\ref{Qratio}) to get equation (\ref{q1l}).
Since the one-loop power spectrum agrees with that in the numerical 
simulations down to scales where the one-loop correction dominates over 
the tree-level contribution, when $\Delta(k_1) > 1$ we use  the full 
expression in equation (\ref{Qratio}), denoted as ``one-loop~(s),'' 
which saturates at large $kR_0$ due to self-similarity (Scoccimarro 1997).
The two bottom panels in Figure~\ref{fig1} illustrate this situation 
for $\Delta(k_1)=1.95$, 3.92, where the simulation results show a 
gentle flattening of $Q(\theta)$ as we approach more non-linear scales.

\begin{figure}[t!]
\centering
\centerline{\epsfxsize=18. truecm \epsfysize=18. truecm
\epsfbox{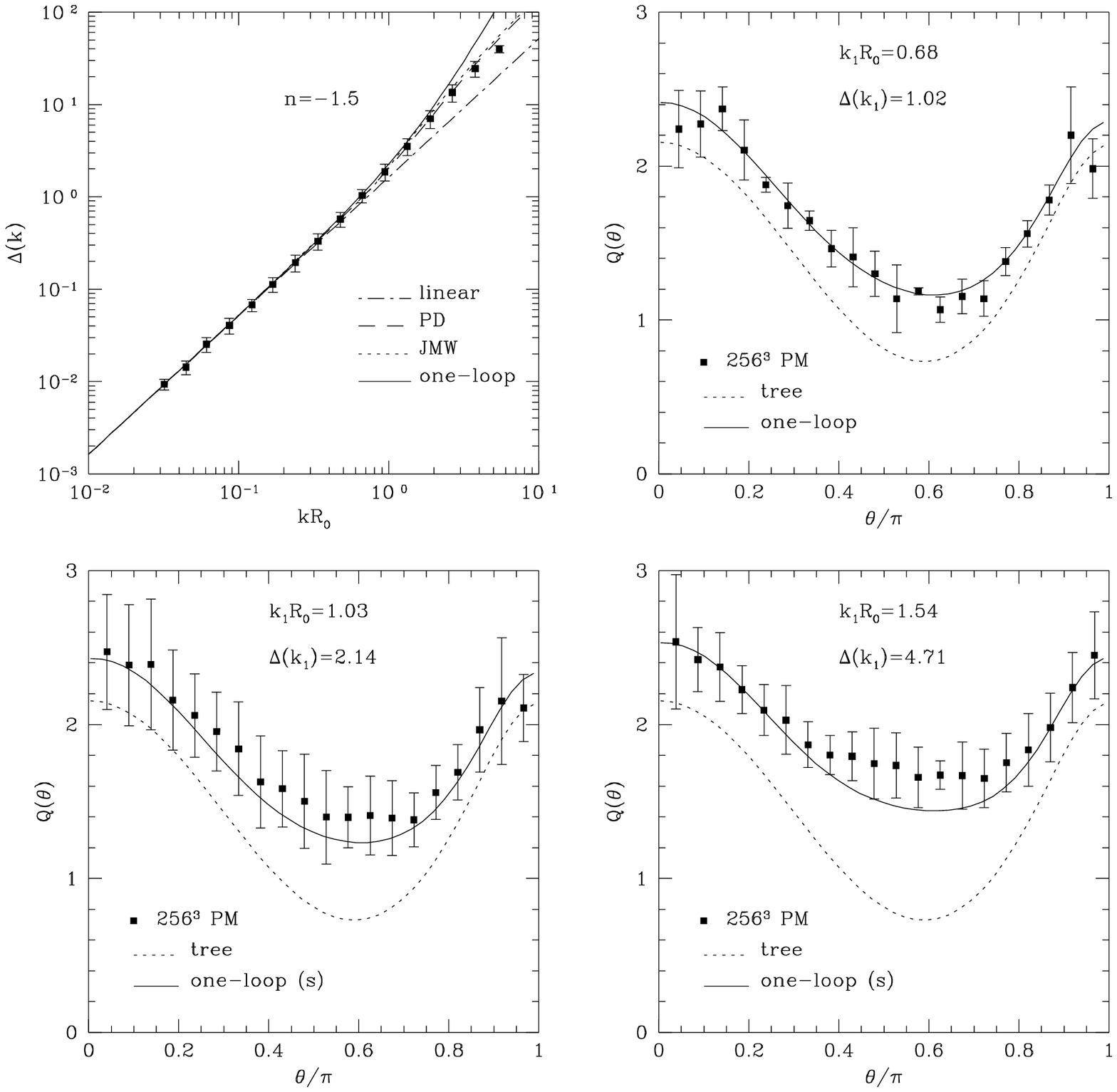}}
\caption{Same as Figure~\protect{\ref{fig1}} for n=-1.5.
}
\label{fig2}
\end{figure}

\begin{figure}[t!]
\centering
\centerline{\epsfxsize=18. truecm \epsfysize=18. truecm
\epsfbox{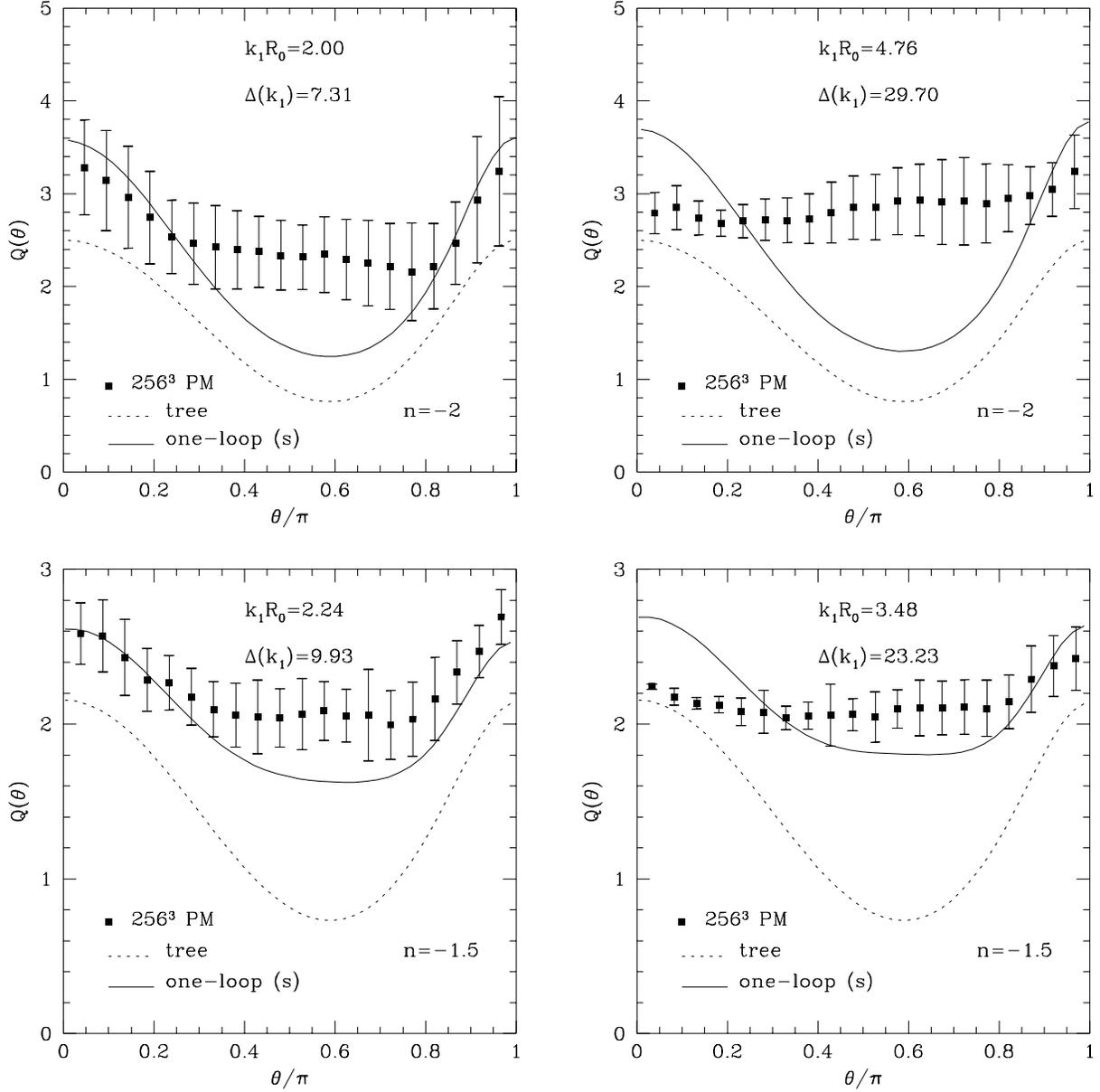}}
\caption{Same as Figure~\protect{\ref{fig1}} and
\protect{\ref{fig2}} for smaller
(more non-linear) scales, showing the decrease in the configuration
dependence  of $Q$. The top panels correspond to the $n=-2$ initial spectrum,
the bottom panels show the $n=-1.5$ initial spectrum.
}
\label{fig3}
\end{figure}

In Figure~\ref{fig2} we present a similar set of plots corresponding
to the $n=-1.5$ spectrum. 
The one-loop power spectrum is given by (Scoccimarro \& Frieman 1996b):
\beq
\Delta(kR_0) = 1.632 \ (kR_0)^{3/2} \
\Big( 1 + 0.391 \ (kR_0)^{3/2}
\Big)
\label{Deltanm1p5},
\eeq
which (see top left panel)
again shows very good agreement with the numerical simulations and
the JMW and PD fitting formulae up to scales where $\Delta(k) \approx 10$.
In the remaining three panels, we show the hierarchical amplitude $Q$ for
stages of non-linear evolution similar to those of the corresponding
panels in Figure~\ref{fig1}.
We see the same trend with scale, namely, a departure from the 
tree-level PT prediction and then a hint of decrease in the configuration 
dependence of $Q$.
Note that, in this case, the one-loop corrections to $Q$ are smaller 
than in the $n=-2$ case, both for the power spectrum and bispectrum.
In Figure~\ref{fig3}, we show $Q$ approaching the strong 
clustering regime, $\Delta(k) \gg 1$, for both $n=-2$ (top panels)
and $n=-1.5$ (bottom panels).
These simulation results confirm the flattening of $Q$ seen in
previous work at small scales (FMS).
Interestingly, the $\theta \simeq 0$ configurations (which correspond
to more non-linear scales) seem to flatten before the $\theta \simeq \pi$
configurations. A similar effect is apparent in Figures~\ref{fig1} 
and \ref{fig2}: the loop corrections to the tree-level result are 
consistently larger for $\theta \simeq 0$ configurations.

We also see from Figures~\ref{fig1}, \ref{fig2}, and
\ref{fig3} that the flattening of $Q$ happens first at
intermediate angles, and then spreads to smaller and larger values 
of $\theta$.
This effect can be interpreted as follows: as explained in \S~2 
(see discussion after eq.~[\ref{q1l}]), on weakly nonlinear scales, 
before shell-crossing, large-scale 
flows are mostly parallel to density gradients, an effect which favors 
collinear configurations ($\theta=0,\pi$).
On smaller, more non-linear scales, the previrialization associated 
with shell-crossing leads to a ``randomization'' of gradients, 
i.e., configurations which do not pick out a preferred direction 
are given relatively more weight.
This helps to explain why the flattening of $Q$ first develops at 
intermediate values of $\theta$.

\begin{figure}[t!]
\centering
\centerline{\epsfxsize=18. truecm \epsfysize=18. truecm
\epsfbox{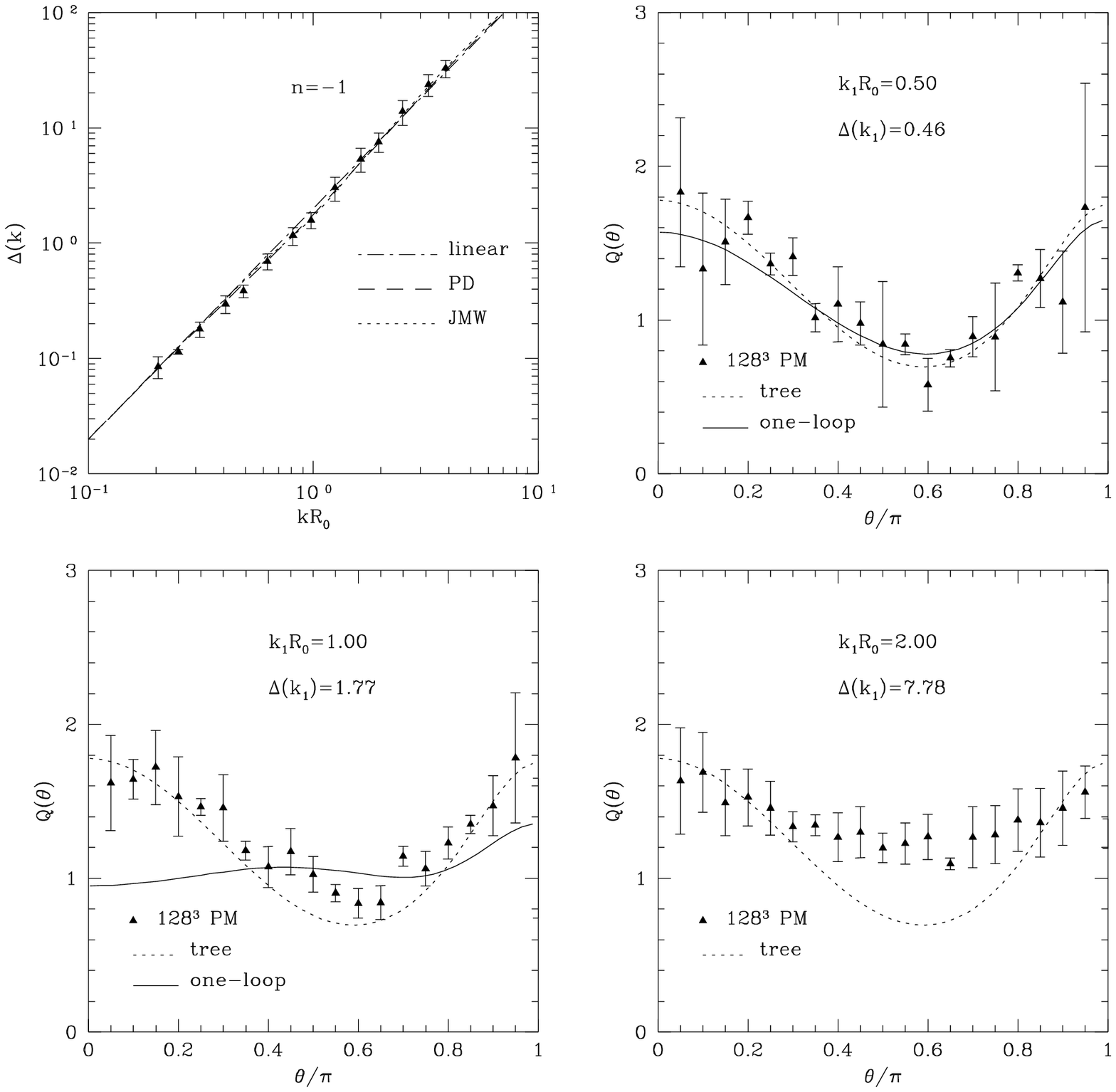}}
\caption{Same as Figure~\protect{\ref{fig1}} for $n=-1$.
}
\label{fig4}
\end{figure}

\begin{figure}[t!]
\centering
\centerline{\epsfxsize=18. truecm \epsfysize=18. truecm
\epsfbox{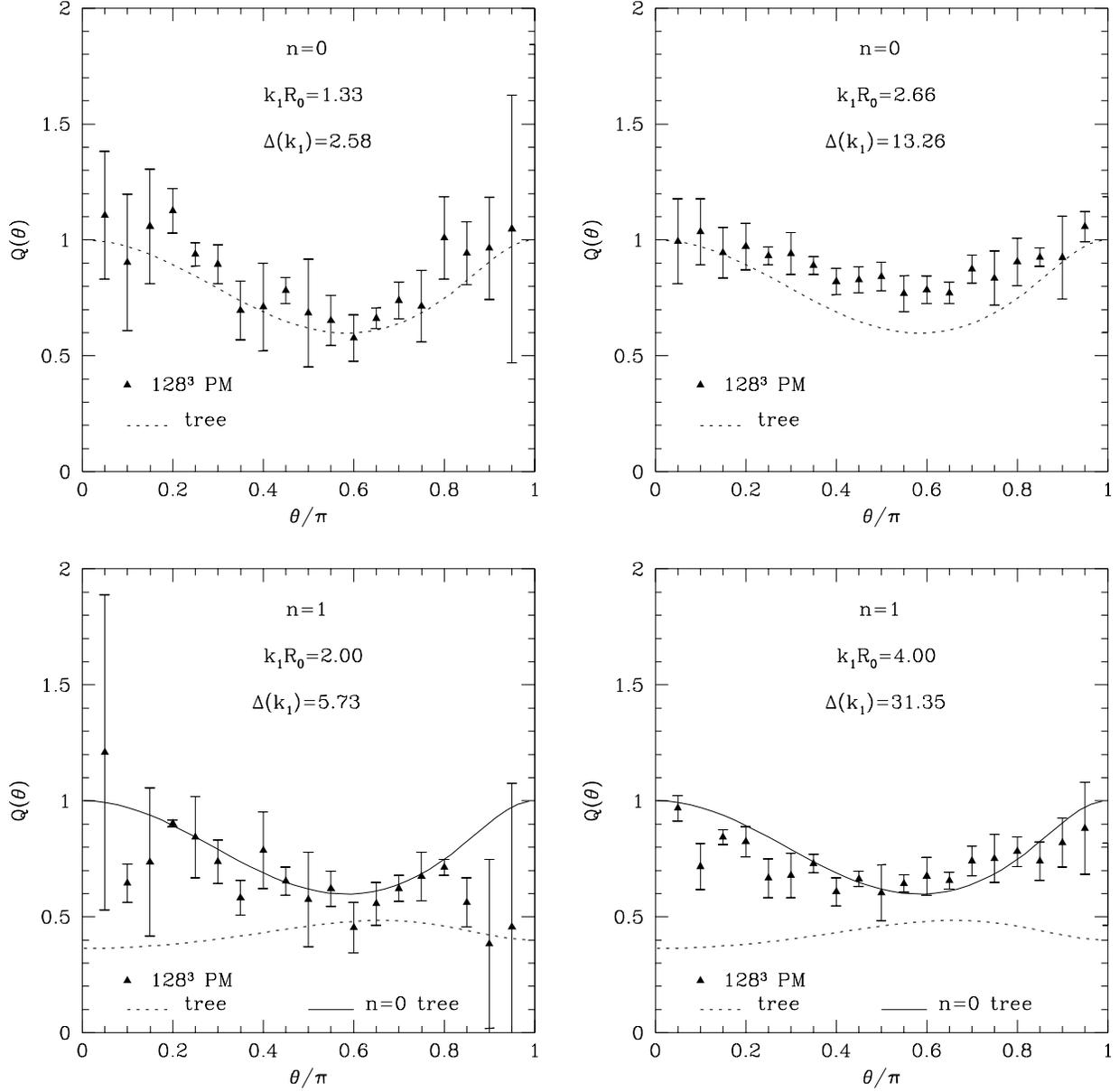}}
\caption{Same as Figure~\protect{\ref{fig1}} for $n=0$ (top panels),
and $n=1$ (bottom panels).
}
\label{fig5}
\end{figure}

Figures~\ref{fig4} and~\ref{fig5} show the results for $n=-1$,
$0$, $1$ spectra from the 128$^3$ PM simulations.
As the spectral index $n$ increases, the relative uncertainties 
in $Q$ increase (see FMS, or Appendix~A.2, eq.~[\ref{eq:qerr}]),
and it becomes more difficult to measure $Q$ accurately in the simulations.
Also, one-loop PT does not work well for these spectra, since ultraviolet
divergences must be regulated by introducing a small scale cutoff, $k_c$, 
which violates self-similarity.
If the one-loop correction to the power spectrum were plotted in the top 
left panel in Figure~\ref{fig4}, we would have different predictions 
for different values of $k_c R_0$ (see Scoccimarro \& Frieman 1996b).
The solid lines in Fig.~\ref{fig4} show that by choosing the small-scale 
PT cutoff as the Nyquist frequency, it is not possible to match the results
of the numerical simulation measurements. 
On the other hand, we see that the non-linear corrections found in the
$N$-body data are very small for $n=-1$, for both the power spectrum and 
$Q$, and tree-level PT does well even on scales for which $\Delta \approx 1$.

A similar situation is shown in the top panels of
Figure~\ref{fig5} for the $n=0$ case.
For $n=1$ (bottom panels of Figure~\ref{fig5}), the simulation data 
are quite noisy, which makes it difficult to reach any conclusion.
Note, however, that the measured $Q$ tends to be systematically above 
the tree-level prediction for $n=+1$, and actually agrees quite well with 
the tree-level predictions for $n=0$ (solid lines).
This might be explained by the following argument.
The initial spectrum of an $n=+1$ simulation is given by 
$P(k)\propto k$ up to the Nyquist frequency of the particles.
For $k>k_{ny}$, however, the initial $N$-body spectrum is
generally white noise, $n_{\rm eff}=0$, or something close to
it, depending on the initial setup of the particle positions 
(e.g., Juszkiewicz et al.~1993; Colombi, Bouchet \& Hernquist~1996).
Usually, the details of the shape of the initial power spectrum at 
these small scales are unimportant, since power is expected to cascade 
from large to small scales, eventually establishing nearly correct small-scale 
behavior regardless of the initial state at such scales 
(e.g., Beacom et al. 1991; Little et al. 1991; Melott \& Shandarin 1993; 
Bagla \& Padmanabhan 1997).
Since an $n=+1$ simulation does not have much relative power at 
large scales, however, the relaxation time for the cascade may be 
long enough for the initial conditions to affect three-point
statistics such as the bispectrum at relatively late output times.
Moreover, this effect may not be manifest in the power spectrum, 
which is insensitive to phase correlations.
In this respect, the analysis of Fry, Melott \& Shandarin (1992) and
FMS suggests that,  
once the system {\it has} relaxed far enough into the nonlinear regime, 
the bispectrum is not significantly affected by the small-scale behavior 
of the initial conditions near the Nyquist frequency, even for $n=+1$.
(Their conclusions are based mainly on the analysis of equilateral 
configurations, however, and it would be interesting to extend this 
study to other configurations.)

\begin{figure}[t!]
\centering
\centerline{\epsfxsize=18. truecm \epsfysize=18. truecm
\epsfbox{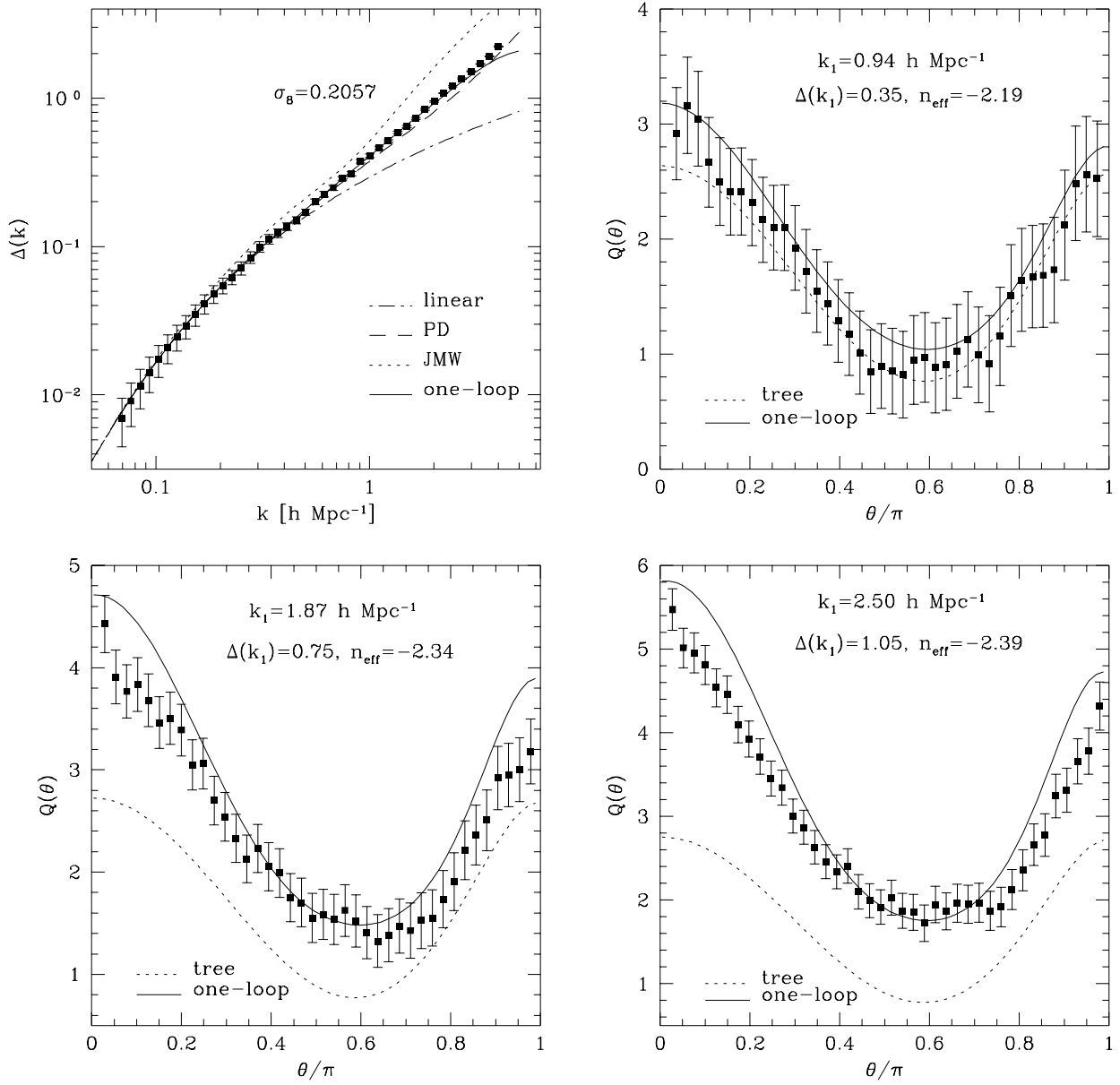}}
\caption{Same as Figure~\protect{\ref{fig1}} for CDM initial
power spectrum, with $\Gamma=0.25$. These four panels correspond
to a $\sigma_8=0.2057$ output.
}
\label{fig6}
\end{figure}

Figure~\ref{fig6} shows the results from the CDM simulation
for $\sigma_8=0.2057$, the earliest output we analyzed
in this case.
For the linear CDM spectrum, we use the BBKS transfer function 
(Bardeen et al.~1986).
For this weakly nonlinear output, where the amplitude of the
power spectrum is not large compared to the white noise level,
we have not corrected for the discrete nature of particles.
Since the initial conditions are set by Zel'dovich displacements
from a grid, Poisson noise is not a good model at early times
(see, e.g., Melott \& Shandarin 1993; Baugh \& Efstathiou 1994).
In fact, the standard shot noise correction would make the non-linear 
power spectrum even smaller than the linear spectrum.
The excellent agreement of the uncorrected $N$-body power spectrum with
one-loop PT and the PD fitting formula seems to indicate that this is
the correct procedure.
For the hierarchical amplitude $Q$, the effect of a full correction for
discreteness would be smaller than on $P(k)$: the corresponding values
of $Q$ in Figure~\ref{fig6} would be somewhat higher than
the results shown, in better agreement with the one-loop calculation
for $\theta=0$ and $\theta=\pi$ (we do not plot this result for reasons
of clarity.) All the other measurements
in this paper have been corrected for discreteness. 
Overall, Figure~\ref{fig6} shows very good agreement between 
one-loop PT and the numerical simulation measurements, 
even when the deviations from the tree-level
PT predictions are dramatic, as can been seen on the bottom right
panel, corresponding to $\Delta(k_1)=1.05$.
The increase in configuration dependence
of $Q$ with $\Delta(k_1)$ is more important than for the $n=-2$ case,
consistent with what one would expect from the discussion in \S~2 and
the effective spectral indices $n_{\rm eff}(k)= d\ln P^{(0)}/d\ln k$
of the scales considered (displayed in Fig.~\ref{fig6}).
Note that the error bars on the plots probably 
underestimate the true errors.
Having access to only one realization, and without the possibility of using
self-similarity (and thus different output times) as a test
on the accuracy of the results because the CDM spectrum is not
scale-free, we have estimated the error bars from the number of
independent modes in $k$-space contributing to a given configuration
(see Appendix A for details.)

Figure~\ref{fig7} shows the set of plots corresponding to
the next output time analyzed in the CDM simulation. We see that the
logarithmic variance contribution $\Delta$ at one-loop agrees very well
with the $N$-body measurements up to $\Delta (k) \approx 6$.
For the hierarchical amplitude $Q$, given that we have only one 
realization, we could only make accurate measurements on scales for 
which $\Delta (k_1) >1$, so we use equation (\ref{Qratio}) for the 
one-loop prediction. 
We see very good agreement between predictions and
measurements for configurations close to collinear ($\theta=0,\pi$),
and a progressive flattening of $Q$
as we move to smaller scales, in agreement with the results for
$n=-2$ and $n=-1.5$.
The latest CDM output, shown in
Figure~\ref{fig8}, illustrates this behavior of $Q$ further 
in  the non-linear regime.
At the smallest scale shown, $ k_1 = 2.50 \,  h \, {\rm Mpc}^{-1} $, 
the configuration dependence of $Q$ is totally washed out. 
For the power spectrum, one-loop PT does remarkably well 
over the whole range of scales considered, remaining within less 
than 50\% of the numerical simulation measurements and the fitting 
formulae up to scales where $\Delta(k) \approx 100$!

Figure~\ref{fig9} shows the equilateral
hierarchical amplitude $Q_{\rm EQ}$ for the $n=-2$, $-1.5$, $-1$
and CDM initial spectra as a function of scale. For the scale-free
$n=-2$, $-1.5$ spectra, the one-loop predictions are those of
equations (\ref{Qeqnm2}) and (\ref{Qeqnm1p5}) respectively.
Note how well these simple formulae describe the behavior of $Q_{\rm EQ}$ 
from the tree-level value to the transition towards the strongly
non-linear plateau.
For the $n=-1$ case, we do not show the one-loop correction, since as 
mentioned above, it does not obey self-similarity and therefore would not
correspond to a single curve in the left bottom panel in
Figure~\ref{fig9}. For the CDM case, we show the one-loop result
for $\sigma_8=0.33$, which agrees very well with the numerical
simulation measurements up to scales where $\Delta(k) \approx 10$ (see
left top panel in Fig.~\ref{fig7}). The $\sigma_8=0.64$ output,
however, corresponds mostly to scales already in the non-linear regime (see
left top panel in Fig.~\ref{fig8}); in this case 
equation (\ref{Qratio}) for equilateral configurations (not shown in
Fig.~\ref{fig9}) underestimates $Q_{\rm EQ}(k)$, as expected from
the discussion above.

An interesting question is whether the hierarchical amplitude $Q$
actually becomes a constant independent of configuration
in the highly non-linear regime.
By comparing the results in Figure~\ref{fig9} with the corresponding
figures for the $k_1/k_2=2$ configurations at the smallest scales, we
see that the equilateral configurations seem to attain a slightly higher
value for $Q$ at large $kR_0$ than the non-equilateral configurations.
Given the uncertainties in our measurements, however, this trend is not
statistically very significant, and more accurate measurements would
be needed to assess in detail the validity of the hierarchical ansatz in the
strongly non-linear regime. In fact, the simulations analyzed in this work
do not probe very deeply into this regime. The validity of the
hierarchical ansatz in the strongly non-linear regime has been
considered in
real space studies of the three and four-point correlation functions and
counts in cells analyses (Bouchet \& Hernquist 1992; Lahav et
al. 1993; Colombi et al. 1994; Lucchin et al. 1994; 
Matsubara \& Suto 1994; Suto \& Matsubara 1994; 
Bonometto et al. 1995; Colombi et al. 1996; Ghigna et al. 1996.) After
correction for finite volume effects, these results seem to show a small
but significant departure from the scaling expected in the
hierarchical ansatz. 

%
%
\section{Conclusions and Discussion}
%
%

We have considered one-loop corrections to the bispectrum in Perturbation
Theory (PT) and
compared the results with numerical simulations for scale-free and CDM initial
spectra with Gaussian initial conditions, focusing on the change
of configuration dependence of the hierarchical amplitude $Q$ as
the transition to the non-linear regime is made. We found very good
agreement between one-loop PT and our $N$-body measurements for scale-free
spectra with $n=-2, -1.5$, and for the CDM initial spectrum. For scale-free
$n \geq -1$ spectra, one-loop corrections diverge, and the simplest
remedy of introducing a cutoff at small scales in the initial
power spectrum (e.g., at the Nyquist frequency of the particles),
breaks self-similarity, an effect which is not seen in
the numerical simulations.
On the other hand, for spectra with $n\geq -1$, tree-level PT does well 
compared to numerical simulations even on scales comparable to the 
correlation length.

For the power spectrum, we find excellent agreement between
one-loop PT and the numerical simulations even on scales where
the one-loop correction dominates over the tree-level contribution, 
i.e., where one would naively expect PT to break down.
In fact, the simple expressions in equations (\ref{Deltanm2}) and (\ref{Deltanm1p5}) 
follow quite closely the fitting formulae for the non-linear power
spectrum proposed by Jain, Mo \& White (1995) and by Peacock
\& Dodds (1996) over a remarkable range of scales.
For the hierarchical amplitude
$Q$, we showed that one-loop corrections correctly describe the
evolution of the configuration dependence observed in numerical simulations
on weakly non-linear scales, for power spectra with sufficient
relative large-scale power ($n <-1$ and CDM).
At scales comparable to the correlation
length, where one-loop contributions become of the same order as
their tree-level counterparts,
the numerical simulations show a progressive flattening of
$Q(\theta)$. This flattening starts at intermediate angles, as these
configurations become increasingly
probable due to ``randomization'' of density and velocity gradients,
and propagates to collinear configurations ($\theta=0,\pi$). One-loop
PT does not reproduce this observed flattening very well, but it is
nonetheless able to follow configurations close to collinear further
into the non-linear regime. In the strong clustering regime,
the $N$-body results
show almost no dependence on configuration shape, $Q \approx$
constant. This saturation
value shows a clear dependence on the initial spectrum, consistent
with previous numerical simulations in the literature (Efstathiou et
al.~1988; Colombi et al.~1996), as parametrized by FMS,
$Q_{sat}(n) \approx 3/(3+n)$.
Note that for the CDM case, a similar formula could
be used, where $n$ is taken as the effective spectral index at the
scale of nonlinearity.
In this case, we would find that, for the
$\sigma_8=0.64$ output time, $n_{\rm eff}(k)
\approx -1.8$ at the scale where $\Delta(k) =1$; this corresponds
to $Q_{sat}(n_{\rm eff}) \approx 2.5$,
in rough agreement with Figures~\ref{fig8}
and \ref{fig9}.

In recent work related to our paper,
Jing \& B\"orner (1996) noted that the predictions
of tree-level PT for the three-point function in real space did not
agree with their numerical simulations of CDM models in the weakly
non-linear regime. For this comparison, however, they considered
scales close to the correlation length, for which one-loop corrections
are expected to significantly alter the tree-level predictions.
The results we obtain in Fourier space suggest that their measurements
should agree very well with PT when one-loop corrections are included.

Finally, in light of these results, we return to comment on the issue 
raised in the introduction, the shape of $Q$ as a probe of bias and its
robustness to non-linear effects.
Although the non-gravitational
effects that transform the non-linear density field into the
observed distribution of luminous galaxies are undoubtedly complex,
they may be relatively local; that is, suppose the probability of
forming a luminous galaxy depends only on the underlying density
field in its immediate vicinity. For simplicity, we also suppose
this dependence is deterministic rather than stochastic. Under
these simplifying assumptions, the relation between the galaxy
density field $\d_g(\x)$ and the mass density field $\d(\x)$ is of the form
$\d_g(\x) = f(\d(\x)) = \Sigma_n b_n \d^n$, where $b_n$ are the
bias parameters.
For the reduced tree-level bispectrum, this local bias scheme implies
\beq
Q_g^{(0)} = {1 \over b_1} Q^{(0)} + {b_2 \over b^2_1}
\label{Qg}
\eeq
(Fry \& Gazta\~{n}aga 1993).
Gazta\~{n}aga \& Frieman (1994) have used the corresponding 
relation for the skewness $S_3$ to infer $b_1 \simeq 1$, 
$b_2 \simeq 0$ from the APM catalog, but the results are 
degenerate due to the relative scale-independence of $S_3$.
Fry (1994b) has used the comparison between $Q_g$ inferred 
from the Lick catalog and the tree-level PT prediction $Q^{(0)}$ 
to infer values for $b_1, b_2$.
In particular, since $Q_g$ displays little of the configuration 
dependence expected of $Q^{(0)}$, he finds a best fit value of 
$b_1 \simeq 3$, a large linear bias factor. On the other hand, in order to 
extract a statistically significant $Q_g(\theta)$ from the Lick catalog,
an average over scales which include values of $k$ comparable to the 
scale of nonlinearity was required.
The question then becomes, to what degree do one-loop
corrections to  $Q^{(0)}$ on these scales affect the determination of
the $b_n$ from equation (\ref{Qg})?

To partially address this question,
in Figure~\ref{fig10} we show the expected correction to the
hierarchical amplitude $Q$ for $k_1/k_2=2$ configurations according to
one-loop PT, for $\Gamma=0.25$ CDM with $\sigma_8=0.64$,
at scales where $\Delta \approx 1$.
Note that we do have $N$-body results on these scales
at this output time, but the statistical uncertainties are rather
large due to the small number of independent modes available (see
Appendix A.) The one-loop PT results in Figure~\ref{fig10} agree
within the errors with these measurements. For a realistic spectrum,
Figure~\ref{fig10} illustrates how the configuration dependence
of $Q$ should change on scales relevant for observations.
There is a minor although noticeable flattening of $Q$; if one were 
to instead attribute this to an effective bias according to equation 
(\ref{Qg}),  it would correspond to 
$1.25 \la b_1 \la 1.4$, $0.5 \la b_2 \la 0.8$, where low (high) 
values of $b_1$ are correlated with low (high) values of $b_2$.
This preliminary result shows that the estimate of the bias parameters
from the measured bispectrum in the galaxy distribution using the
tree-level result, equation (\ref{Qg}), could be affected by
non-linear evolution, unless scales much larger than the scale of 
non-linearity are considered. Further work is needed in order to 
quantify this issue better.

In evaluating the prospects for measuring the bispectrum in galaxy surveys,
observational considerations such as selection function, angular
projection in two-dimensional surveys, redshift distortions,
and aspects of survey design such as sky coverage, geometry, and
sampling rate must be carefully considered.
Although an exhaustive study of these kinds of questions is beyond
the scope of this paper, we shall make some general comments.

On large scales, for a scale-free power spectrum, $ P(k) \sim k^n $, the
statistical uncertainty in $Q$ scales with configuration size
as $ (\knl/k)^{(3+n)/2} $ per mode (FMS).
To compare observations with perturbation theory, we need data on
scales $ k < \knl $ and thus need a survey with many modes at the
scale $ \knl $, i.e., a survey that covers a large volume.
In order to determine the optimal sampling strategy, it is necessary to
know with precision what are the various sources of error, at least from
the pure statistical point of view (e.g., Szapudi \& Colombi 1996).
To reduce shot-noise uncertainties, it is desirable to construct surveys
with a large number of galaxies, i.e., as complete as possible.
In addition, minimizing edge effects requires compactness (i.e., the
boundaries of the survey should have minimal surface), while
finite-volume errors call for large sky coverage.
The best sampling strategy results from balancing these
various effects.
Kaiser (1986) concluded that to measure the two-point function at
large scales it is best to have sparse samples with large sky coverage
and a sampling rate of order 1/10.
In more recent work concerning the power spectrum,
Heavens \& Taylor (1997) reach similar conclusions.
In the case where only a small part of the sky is covered, another
issue arises: the choice of the catalog geometry.
The conclusions of Kaiser (1996), who analyzed weak gravitational
lensing statistics,
favor a catalog made of many small patches spread over the sky.
A similar study by Colombi, Szapudi, \& Szalay (1997), based on
counts-in-cells statistics (including higher order moments),
reaches at least qualitatively the same conclusions;
however, the latter conclude that the sampling rate should be
increased as higher order statistics or smaller scales are considered.

The discussion above can be illustrated by the situation in existing
surveys, where there are on-going efforts to measure
three-point statistics.
In the QDOT and IRAS 1.2Jy surveys,
the main limitation comes from discreteness effects,
which dominate the signal even at large scales (Feldman et al. 1997).
On the other hand, in the APM survey (Gazta\~naga \& Frieman 1997),
the main challenge is to successfully deconvolve the angular projection,
to extract the signal at quasilinear scales (Thomas \& Fry
1997). The situation in the Las Campanas Redshift Survey
is somewhat complicated, because of both
geometry and selection function effects.
In particular, the thin slices make the estimation of three-dimensional
statistics rather uncertain (see, e.g., Heavens \& Taylor 1997),
even for the power spectrum case.
For the bispectrum, the mixing of scales arising from the window
function of a thin slice makes it at best difficult to separate out the
quasilinear regime and compare with perturbation theory calculations.
The prospects are much better for planned future surveys; in particular,
the Sloan Digital Sky Survey and the Two Degree Field
Survey should provide an accurate determination
of the bispectrum over
a wide range of scales in the weakly non-linear regime, with errors
perhaps as small as a few percent in $Q$ (Colombi, Szapudi, \&
Szalay 1997).

Galaxy clustering derived from redshift surveys is distorted
radially by peculiar velocities (Kaiser 1987).
An important issue regarding the determination of bias using
equation (\ref{Qg}) in redshift surveys is how redshift distortions
are expected to modify the configuration dependence of $Q$.
At large scales, tree-level PT predicts that redshift distortions
increase the configuration dependence of $Q^{(0)}$ for models with
$\Omega=1$ by 20\%, whereas low $\Omega$ models show negligible redshift
distortions (Hivon et al. 1995).
This is as expected: peculiar velocities from coherent inflows, most
important in the $\Omega=1$ case, lead to more anisotropic structures
in redshift space, thus increasing the configuration dependence of $Q$.
On small scales, on the other hand, velocity dispersion makes $Q_{EQ}$
less scale-dependent than in real space, as shown by modelling the
velocity distribution function (Matsubara 1994) and numerical simulations
(Lahav et al. 1993; Matsubara \& Suto 1994; Suto \& Matsubara 1994;
Bonometto et al. 1995; Ghigna et al. 1996).
Although these works have concluded that higher-order statistics
are therefore more hierarchical in redshift space than in real space, analysis
of non-equilateral configurations leads in fact to the opposite
conclusion: at small scales, the redshift space
bispectrum shows increased configuration
dependence due to anisotropies caused by velocity correlations along the
line of sight, which enhance colinear configurations and suppress
equilateral configurations relative to the real space case
(Scoccimarro, Couchman \& Frieman 1997).

An attractive feature of equation (\ref{Qg}) as a tool to probe bias, is that 
the tree-level hierarchical amplitude $Q^{(0)}$ is very
insensitive to the cosmological parameters $\Omega$ and $\Lambda$, in 
contrast with determinations from large-scale flows which contain a 
degeneracy of the linear bias parameter $b_{1}$ with $\Omega$. It is
interesting to check whether one-loop corrections to the bispectrum
introduce a  
significant dependence on cosmological parameters. In Appendix B.3 we 
show that to a very good approximation, all the dependence of PT 
solutions on $\Omega$ 
and $\Lambda$ can be described by the linear growth factor, to 
arbitrary order in PT. 
Therefore, for the same normalization of the linear power spectrum, 
or $\sigma_{8}$, the hierarchical amplitude $Q$ should be almost 
insensitive to  $\Omega$ and $\Lambda$ even in the  
non-linear regime. We have checked this prediction against numerical 
simulations, and found that for the $\sigma_{8}=0.64$ output, the 
reduced bispectrum $Q$ in an $\Omega=0.5$ ($\Gamma=0.25$) open CDM model 
is virtually indistinguishable from the corresponding plots 
shown in Figure~\ref{fig8}. The results in Appendix B.3 also 
suggest that the same result regarding the $\Omega$ and $\Lambda$  
dependence should hold for higher order statistics, such 
as the $S_{p}$ parameters ($p=3,4,\ldots$). Work is on progress on 
this issue (Jain \& Colombi 1997). 

\acknowledgments
After this work was completed, we received a preprint by 
Matarrese, Verde, and Heavens (1997), which discusses error estimates on
bispectrum measurements and a likelihood approach to extracting
the bias.

We would like to thank F.~Bernardeau, E. Gazta\~{n}aga, B.~Jain, 
R.~Juszkiewicz, and C.~Murali for conversations and especially 
H. Couchman and D. Pogosyan for numerous helpful discussions.
Parts of this work were done while J.N.F and J.A.F. were visitors at
the Aspen Center for Physics, and while E.H. was at Institut
d'Astrophysique de Paris, supported by the Minist\`ere de la Recherche
et de la Technologie.
This research was supported in part by the DOE at Chicago and
Fermilab and by NASA grants NAG5-2788 at Fermilab and NAG5-2835
at the University of Florida. Research at the University of Kansas was
supported by the NSF-EPSCoR program and NASA grant NAGW-3832. 
The CDM simulations analyzed in this work were obtained
from the data bank of cosmological $N$-body simulations provided by the
Hydra consortium ({\sf http://coho.astro.uwo.ca/pub/data.html}) and produced
using the Hydra $N$-body code (Couchman, Thomas, \& Pearce 1995).
The 128$^3$ PM simulations were produced under a grant from the
National Center for Supercomputing Applications, Urbana, Illinois.
The computational means (CRAY-98) to do the PM simulations with $256^3$ 
particles were made available to us thanks to the scientific council of 
the Institut du D\'eveloppement et des Ressources en Informatique 
Scientifique (IDRIS).

\appendix
\section{Measuring the Bispectrum in Numerical Simulations}

\subsection{The algorithm}

To measure the power spectrum and the bispectrum
in the $256^3$ scale-free simulations
and in the CDM simulation, we wrote a FORTRAN program which, in brief,
computes the density contrast on a grid by using
``cloud-in-cell'' (CIC) interpolation (see, e.g., Hockney \& Eastwood 1988),
fast Fourier transforms it, and then uses Monte-Carlo simulations for
spatial averaging in $k$-space.

More specifically, given a triangle with
sides $k_1$, $k_2$, and $k_3$, the estimates of $P(k_1)$, $P(k_2)$,
$P(k_3)$ and $B(k_1,k_2,k_3)$ are done as follows. The quantities
we measure are actually smoothed over a bin of width $\Delta k$:
\beq
 {\hat P}(k)=\frac{1}{V(k)}
 \int_{k-\Delta k/2}^{k+\Delta k/2}  \breve{P}(q) \ q^2 dq,
\eeq
with
\beqa
 \breve{P}(q) &=&
 \int   {\hat \delta}(\q) \
 {\hat \delta}^*(\q) \ \sin\theta \ d \theta \ d \phi, \\
  V(k) &\equiv&  4\pi k^2\Delta k [1+\Delta k^2/(12k^2)],
  \label{eq:vk}
\eeqa
and
\beq
 {\hat B}(k_1,k_2,k_3) = \frac{1}{V(k_1,k_2,k_3)}
	\int_{q_i \in [k_i-{\Delta k}_i/2, k_i + {\Delta k}_i/2]}
         \breve{B}(q_1,q_2,q_3) \ q_1^2 dq_1 \ q_2^2 dq_2 \
        q_3^2 dq_3,
	\label{eq:smoothB} 
\eeq
with
\beqa
  \breve{B}(q_1,q_2,q_3) &=&
 \int_{\q_1+\q_2+\q_3=0}
 {\rm Re}\left[ {\hat \delta}(\q_1) {\hat \delta}(\q_2)
 {\hat \delta}(\q_3) \right]
 \prod_{i=1}^{3} \left (\sin\theta_i d \theta_i d \phi_i \right),
	\label{eq:calcB} \\
  V(k_1,k_2,k_3) &=& \int_{q_i \in [k_i-{\Delta k}_i/2, k_i + {\Delta
  k}_i/2]}  [\dD] \; d^3q_1 d^3q_2 d^3q_3=8\pi^2 k_1 k_2 k_3
  \Delta k_1 \Delta k_2 \Delta k_3.
  \label{eq:vk123}
\eeqa
In equation (\ref{eq:calcB}), ``Re'' means ``real part of''. 
Most importantly, the function ${\hat \delta}(\q)$ is corrected for
the effects of the smoothing in real space due to CIC interpolation,
yielding
\beq
   {\hat \delta}(\q)=\frac{k_x \, k_y \, k_z} 
   {8 \sin( k_x/2)\sin( k_y/2)\sin( k_z/2)}  {\tilde \delta}(\q),
\eeq
where ${\tilde \delta}(\q)$ is the actual Fourier
transform of the density contrast computed on the grid.

There is a constraint on the values of $k_1$, $k_2$, and $k_3$ so
that the numbers $q_i \in {\cal D}_i$, where
\beq
   {\cal D}_i=[k_i-{\Delta k}_i/2, k_i+{\Delta k}_i/2],
\eeq
form a triangle, that is
\beq
   k_3 \geq \left| k_1-k_2 \right| + 3\Delta k/2
  \quad \hbox{and cyclic permutations,}
  \label{eq:contrainte}
\eeq
with
\beq
   \Delta k=(\Delta k_1 + \Delta k_2 + \Delta k_3)/3.
\eeq
As a result, if $\theta$ represents the angle between vectors $\k_1$
and $\k_2$, the sampled values of $\theta$ cannot be arbitrarily close to
0 or close to $\pi$: 
\beq
   \left[ \cos\theta \right]_{\rm min} = -1
   + \frac{ 3 \Delta k (k_1+k_2) /2 - (3 \Delta k)^2/8 }{k_1 k_2}~,
\eeq
and
\beq
   \left[ \cos\theta \right]_{\rm max} = 1-
   \frac{3 \Delta k \left| k_1 - k_2 \right|/2 + (3 \Delta k)^2/8}{k_1 k_2}~.
\eeq
Now, let us imagine that we have chosen numbers $k_1$, $k_2$, and
$k_3$ obeying the constraint of equation (\ref{eq:contrainte}).
Calculating integral (\ref{eq:smoothB}) is simply done by Monte-Carlo 
simulation, randomly choosing numbers $q_i$ in the intervals of equal 
probability ${\cal D}_i$.
We simultaneously compute integral (\ref{eq:calcB}), i.e., 
we estimate the average of the quantity
${\rm Re}[ \delta(\q_1) \delta(\q_2) \delta(\q_3) ]$
over all the possible positions of the solid body formed by the vectors 
$q_1$, $q_2$, and $q_3$.
The method used here consists of picking a random
direction for $\q_1$ and then choosing randomly the direction of
$\q_2$ in a circle of equal probability around $\q_1$ such
that the angle between $\q_2$ and $\q_1$ remains fixed
(actually determined by the values of $q_1$, $q_2$, and $q_3$).
Each iteration in our Monte-Carlo simulation thus consists
of randomly choosing the numbers $q_1$, $q_2$, $q_3$
and the orientation of the solid formed by the vectors $\q_1$,
$\q_2$, and $\q_3$ in space (three angles).

There is the problem that we have access to only discrete values of
$\q$ on a three-dimensional grid $\q_{i,j,k}$. This is solved by ``random
interpolation''. For a given value of $\q$, we associate a
probability to each of the eight nearest grid sites.
These weights are computed the same way as for CIC interpolation. Each
of these weights determines the probability of actually picking the
corresponding value of $\q_{i,j,k}$, i.e., setting
$\q\equiv \q_{i,j,k}$. To measure the bispectrum we apply this
procedure to $\q_1$ and $\q_2$ and set $\q_3=-\q_1-\q_2$.

Finally, one might wish to correct for discreteness effects, that is, to
subtract off the contribution of the shot noise of the particles
(see, e.g., Peebles 1980, eqs.~[41.5] and [43.6]).
We did so for all the measurements, except for the earliest output of
the CDM simulation we analyzed, as explained in \S~5.

\subsection{Error Bars}

The method of estimating the errors in the figures depends on the case
considered. For the scale-free PM simulations with $128^3$ particles
($n=-2$, $-1$, $0$, $+1$), since we have $N_{\rm rea}=4$ independent
realizations we can infer the errors from the dispersion of the
measurements.
If $F$ is a quantity we measure, for which the estimator ${\hat F}$ is
\beq
  {\hat F}=\frac{1}{N_{\rm rea}} \sum_{i=1}^{N_{\rm rea}} {\hat F}_i,
  \label{eq:est}
\eeq
where ${\hat F}_i$ stands for the measurement of $F$ in realization
$1\leq i \leq N_{\rm rea}$, the estimator of the error reads
\beq
   [\Delta F]^2 = \frac{1}{N_{\rm rea}-1} \sum_{i=1}^{N_{\rm rea}}
   ({\hat F}_i-{\hat F})^2.
   \label{eq:disp}
\eeq

For each of the two scale-free PM simulations with $256^3$ particles, 
we had only a single realization to work with.
In these cases, we extracted results at several output times and 
rescaled them under the assumption of self-similar evolution, 
forming the average of equation (\ref{eq:est}).
To estimate the errors, we treated the different output times as 
effectively different realizations and used the estimator (\ref{eq:disp}).
However, the different output times are not actually statistically 
independent realizations.
In these cases, the error bars on the figures are likely to be more a 
reflection of departures from self-similarity than of real underlying 
statistical uncertainties.

For the CDM simulation, we have access to only one realization, and
we cannot combine rescaled output times
to artificially increase $N_{\rm rea}$, because CDM is not scale-free.
In this case, we use the error estimates of  FMS and
Feldman, Kaiser \& Peacock (1994), which assume
that the Fourier components are Gaussian-distributed: 
\beq
   \left[ {\Delta P(k)}\right]^2 =\frac{1}{{\hat V}(k)}
   \left[  P_{\rm tot}(k) \right]^2,
	\label{eq:err1}
\eeq
\beq
   \left[ \Delta B(k_1,k_2,k_3) \right]^2 =\frac{1}{2{\hat V}(k_1,k_2,k_3)}
   \left[ P_{\rm tot}(k_1) P_{\rm tot}(k_2) P_{\rm tot}(k_3) \right],
	\label{eq:err2}
\eeq
where $P_{\rm tot}(k)=P(k)+1/N_{\rm par}$ is the total power.
(We express wave numbers in units of the fundamental mode 
$k_f\equiv 2\pi/L$.)
Although the power spectrum and the bispectrum are statistically correlated, 
we use the standard error propagation formula to compute the error on
the ratio $Q(k_1,k_2,k_3)$.

In equations (\ref{eq:err1}) and (\ref{eq:err2}), the quantities
${\hat V}(k)$ and ${\hat V}(k_1,k_2,k_3)$ represent, in units of $k_f$,
the ``number of independent modes'' for the power spectrum and the
bispectrum. In principle, they should be equal to $V(k)$ and
$V(k_1,k_2,k_3)$, but this does not make sense when the bins
$\Delta k$ and $\Delta k_i$ are order of unity or smaller (in units of
$k_f$), as is the case for our measurements (we chose $\Delta
k=\Delta k_i=0.01$).
We are indeed working in a discrete Fourier space, in which the
thinnest effective binning is $\Delta k_{\rm eff} \sim 1$.
For $\Delta k$, $\Delta k_i \la 1$,
we therefore take ${\hat V}(k) \sim 2 \pi k^2$ 
and ${\hat V}(k_1,k_2,k_3) \sim (4/3) \pi^2 k_1 k_2 k_3$, where we
included symmetry factors of 1/2 and 1/6 respectively (compare with 
Eqs.~[\ref{eq:vk}] and [\ref{eq:vk123}]). The reality
constraint $\d(\k)^*=\d(-\k)$ reduces the effective Fourier volume by
one-half, whereas the three-fold
symmetry of triangle configurations yields an additional factor of 1/3
(see FMS).  

In the $128^3$ scale-free simulations, for which we have several
independent realizations, we found that the errors one would estimate from
equations (\ref{eq:err1}) and (\ref{eq:err2}) are slightly smaller than 
the dispersion in equation (\ref{eq:disp}).
This result suggests that equations (\ref{eq:err1}) and (\ref{eq:err2}) 
underestimate the true errors. 
We do not know whether this is because of the Gaussian
assumption quoted above or because the number of independent
modes is overestimated. 
Rigorous calculation of the power spectrum and bispectrum error bars is
a non-trivial issue that goes beyond the scope of this paper.

In any case, neglecting the shot-noise contribution, $P_{\rm tot}(k)
\simeq P(k)$, and considering equilateral configurations
$k_1=k_2=k_3$, with the above assumptions one finds
\beq
   \frac{\Delta Q}{Q} \simeq \frac{1}{\sqrt{6\pi\Delta(k)}\ Q}.
	\label{eq:qerr}
\eeq
This means that the relative error on $Q$ is expected to be
larger on large length scales, which is unfortunate 
if one wants to probe the weakly nonlinear regime.
Also, since one expects $Q$ to decrease with spectral index $n$,
the relative error on $Q$ is expected to increase with $n$ for the
same degree of nonlinearity $\Delta(k)$.

Note finally that there is an error due to the finite number 
$N_{\rm iter}$ of iterations used for the Monte-Carlo simulation 
discussed in the previous Section.
A fair estimate of this error to first order is to use equations 
(\ref{eq:err1}) and (\ref{eq:err2}) with ${\hat V}(k)=N_{\rm iter}$ 
or ${\hat V}(k_1,k_2,k_3)=N_{\rm iter}$.
With our choice, $N_{\rm iter}=10^7$, the corresponding uncertainty 
on the measurement of $P(k)$ and $B(k_1,k_2,k_3)$ is negligible compared 
to the other sources of error mentioned above.

\section{Eulerian Perturbation Theory}
\subsection{The Equations of Motion}
\label{eom}

Assuming the universe is dominated by pressureless dust 
(e.g., cold dark matter), in the single-stream approximation 
(prior to orbit crossing) one can adopt a fluid description of the 
cosmological $N$-body problem. 
In this limit, the relevant equations of motion in the Newtonian 
approximation to General Relativity correspond to conservation of mass 
and momentum and the Poisson equation 
(e.g., Peebles 1980, Scoccimarro \& Frieman 1996).
Assuming the initial velocity field is irrotational,
the system can be described completely in terms of the density field and
the velocity divergence, $ \theta \equiv \nabla \cdot \v $.
Defining the conformal time $\tau=\int dt/a$ and the conformal expansion
rate $ {\cal H}\equiv {d\ln a /{d\tau}}$, the equations of motion in
Fourier space become
\beqa
  &&{\partial \tilde{\delta}(\k,\tau) \over {\partial \tau}} +
	\tilde{\theta}(\k,\tau) = - \int d^3 k_1 \int d^3 k_2 
	\dD(\k - \k_1 - \k_2) \alpha(\k, \k_1) \tilde{\theta}(\k_1,\tau) 
	\tilde{\delta}(\k_2,\tau)   \label{ddtdelta}, \\
  &&{\partial \tilde{\theta}({\k},\tau) \over{\partial \tau}} +
	{\cal H}(\tau)\, \tilde{\theta}(\k,\tau) + {3\over 2}  
	\Omega {\cal H}^2(\tau) \tilde{\delta}(\k,\tau) = \nonumber \\
  && \qquad - \int d^3k_1 \int d^3k_2 \dD(\k
	- \k_1 - \k_2) \beta(\k, \k_1, \k_2)
	\tilde{\theta}(\k_1,\tau) \tilde{\theta}(\k_2,\tau)
	\label{ddttheta},
\eeqa
where $\k$ is a comoving wave number, and
\beq
	\alpha(\k, \k_1) \equiv {\k \cdot \k_1 \over{k_1^2}}, 
	\qquad \beta(\k, \k_1, \k_2) \equiv 
	{k^2 (\k_1 \cdot \k_2 )\over{2 k_1^2 k_2^2}}  \label{albe}.
\eeq
Equations (\ref{ddtdelta}) and (\ref{ddttheta}) are valid in an
arbitrary homogeneous and isotropic universe, which evolves according
to the Friedmann equations:
\beqa
  &&{\partial {\cal H}(\tau) \over {\partial \tau}} = - \frac{\Omega}{2}
	{\cal H}^2(\tau) + \frac{\Lambda}{3}a^2(\tau) \label{hdot}, \\
  &&(\Omega-1) {\cal H}^2(\tau) = k - \frac{\Lambda}{3}a^2(\tau)
  \label{keq},
\eeqa
where $\Lambda$ is the cosmological constant, the spatial
curvature constant $k=-1,0,1$ for $\Omega_{\rm tot}<1$, $\Omega_{\rm tot}=1$,
and $\Omega_{\rm tot}>1$, respectively, and 
$\Omega_{\rm tot}\equiv\Omega+\Omega_\Lambda$, with $\Omega_\Lambda
\equiv \Lambda a^2/(3 {\cal H}^2)$.

\subsection{Perturbation Theory Solutions for $\Omega=1$ and $\Lambda=0$}

For $\Omega=1$, the perturbative growing mode solution for $\delta$ is 
given by equation (\ref{ptansatz}) and for the velocity divergence by
\beq
	 \tilde{\theta}({\k},\tau) =
	{\cal H}(\tau) \sum_{n=1}^{\infty} a^n(\tau) \theta_n({\k})
	\label{ptansatz2}.
\eeq
The $n$th order solution for $\delta$ is given by equation 
(\ref{ec:deltan}), with a similar relation for the velocity field,
\beq
	\theta_n({\k}) = - \int d^3q_1 \ldots \int d^3q_n 
	\dD(\k - \q_1 - \ldots - \q_n)  G_n^{(s)}(\q_1, \ldots , \q_n)
	\delta_1(\q_1) \ldots \delta_1(\q_n)   \label{ec:thetan}.
\eeq

The functions $F_n^{(s)}$ and $G_n^{(s)}$ are constructed from the 
fundamental mode coupling functions $\alpha({\k}, {\k}_1)$ 
and $\beta({\k}, {\k}_1, {\k}_2)$ by a recursive procedure 
(see Goroff et al. 1986; Jain \& Bertschinger 1994), 
\beqa
F_n(\q_1,  \ldots , \q_n) &=& \sum_{m=1}^{n-1}
  { G_m(\q_1,  \ldots , \q_m) \over{(2n+3)(n-1)}} \Bigl[ (2n+1) 
  \alpha(\k,\k_1)  F_{n-m}(\q_{m+1}, \ldots , \q_n) \nonumber \\
&& \qquad + 2 \beta(\k,\k_1, \k_2) 
  G_{n-m}(\q_{m+1}, \ldots , \q_n) \Bigr] \label{ec:Fn}, \\
G_n(\q_1,  \ldots , \q_n) &=& \sum_{m=1}^{n-1}
 { G_m(\q_1, \ldots , \q_m) \over{(2n+3)(n-1)}}
 \Bigl[3 \alpha(\k,\k_1)  F_{n-m}(\q_{m+1}, \ldots , \q_n) \nonumber \\
&& \qquad + 2n \beta(\k, \k_1, \k_2)  
  G_{n-m}(\q_{m+1}, \ldots ,\q_n) \Bigr] \label{ec:Gn}
\eeqa
(where $ \k_1 \equiv \q_1 + \ldots + \q_m$, 
$\k_2 \equiv \q_{m+1} + \ldots + \q_n$,  
$\k \equiv \k_1 +\k_2 $, and $F_1 = G_1 \equiv 1$), 
and the symmetrization procedure:
\label{ec:symm}
\beqa
 F_n^{(s)}(\q_1, \ldots , \q_n) &=& {1\over{n!}} \sum_{\pi }
 F_n(\q_{\pi (1)}, \ldots , \q_{\pi (n)})   \label{ec:Fns}, \\
 G_n^{(s)}(\q_1,  \ldots , \q_n) &=& {1\over{n!}} \sum_{\pi }
  G_n(\q_{\pi (1)},  \ldots   ,\q_{\pi (n)})  \label{ec:Gns}, 
\eeqa
where the sum is taken over all the permutations $\pi $ of the set
$\{1, \ldots ,n\}$.
Explicit expressions for the unsymmetrized kernels
$F_3$ and $F_4$ are given in Goroff et al. (1986).

\subsection{$\Omega$ and $\Lambda$ dependence of Perturbation Theory Kernels}

\def\t{\theta}

When $\Omega \neq 1$ and/or $\Lambda \neq 0$, the PT solutions at each
order become increasingly more complicated, due to the fact that
growing modes at order $n$ in PT do not scale as $a^n(\tau)$ as
assumed in equations (\ref{ptansatz}) and (\ref{ptansatz2}).
Furthermore,
the solutions at each order become non-separable functions of $\tau$
and $\k$ (Bouchet et al. 1992, 1995; Bernardeau 1994b; Catelan et al. 
1995), so there appear to be no general recursion relations for the
PT kernels in an arbitrary FRW cosmology.
However, it is well known that the $\Omega$ and $\Lambda$ dependence 
is extremely weak once the growth factors have been scaled out 
(Bouchet et al. 1992, 1995; Bernardeau 1994b).
In this appendix we show that a
simple approximation to the equations of motion leads to 
separable solutions to arbitrary order in PT and the same recursion
relations as in the Einstein-de Sitter case.
All the information on the
dependence of the PT solutions on the cosmological parameters $\Omega$
and $\Lambda$ is then encoded in the linear growth factor, 
$D_1(\tau)$, which in turn corresponds to the normalization of the linear
power spectrum, or $\sigma_8$.

In linear PT, the solution to the equations of motion 
(\ref{ddtdelta}) and (\ref{ddttheta}) reads
\beqa
\delta(\k,\tau) &=& D_1(\tau) \delta_1(\k), \\
\theta(\k,\tau) &=& - {\cal H}(\tau) f(\Omega,\Lambda) D_1(\tau) 
\delta_1(\k),
\eeqa 
where $D_1(\tau)$ is linear growing mode, which from the
equations of motion must satisfy
\beq
\frac{d^2D_1}{d\tau^2} + {\cal H}(\tau) \ \frac{dD_1}{d\tau} =
\frac{3}{2} \Omega {\cal H}^2(\tau) D_1,
\eeq
and $f(\Omega,\Lambda)$ is defined as 
\beq
f(\Omega,\Lambda) \equiv \frac{d \ln D_1}{d \ln a} = \frac{1}{\cal H}
\frac{d \ln D_1}{d \tau}.
\eeq
Explicit expressions or $D_1(\tau)$ and $f(\Omega,\Lambda)$ are
not needed for our purposes (see e.g. Peebles 1980).
It is nevertheless important to note the following simple fits 
in the cosmologically interesting cases, namely 
$f(\Omega,\Lambda) \approx \Omega^{3/5}$ when $\Lambda=0$ (Peebles 1980), 
and  $f(\Omega,\Lambda) \approx \Omega^{5/9}$ when
$\Omega+\Omega_\Lambda=1$ (Bouchet et al. 1995).
As mentioned before, we look for separable solutions of the form 
\beqa
\delta(\k,\tau) &=& \sum_{n=1}^{\infty} D_n(\tau) \delta_n(\k), \\
\theta(\k,\tau) &=& {\cal H}(\tau) f(\Omega,\Lambda) 
\sum_{n=1}^{\infty} E_n(\tau) \theta_n(\k), 
\eeqa

From the equations of motion (\ref{ddtdelta}) and
(\ref{ddttheta}) we get for the $n^{\rm th}$ order solutions,
\beq
\frac{\dot D_n}{{\cal H} f} \d_n + E_n \t_n = -\int d^3k_1 d^3k_2 
\delta_D(\k-\k_1-\k_2) \alpha(\k,\k_1) \sum_{m=1}^{n-1} D_{n-m}E_m 
\t_m(\k_1) \d_{n-m}(\k_2),
\label{dn}
\eeq

\begin{eqnarray}
\frac{\dot E_n}{{\cal H} f} \t_n & + & \Big(\frac{3\ \Omega}{2f^2}-1\Big) 
E_n \t_n  + \frac{3\ \Omega}{2f^2} D_n \d_n  =  \nonumber \\
& & -\int d^3k_1 d^3k_2 
\delta_D(\k-\k_1-\k_2) \beta(\k,\k_1,\k_2) \sum_{m=1}^{n-1} E_{n-m}E_m 
\t_m(\k_1) \t_{n-m}(\k_2),
\label{tn}
\end{eqnarray}
where the dot denotes a derivative with respect to $\tau$.
By simple inspection, we see that if $f(\Omega,\Lambda) = \Omega^{1/2}$, 
then the system of equations becomes indeed separable, with
$D_n = E_n = (D_1)^n$.
In fact, the recursion relations then reduce to the standard 
$\Omega=1$, $\Lambda=0$ case, shown in equations (\ref{ec:Fn})
and (\ref{ec:Gn}).
Then $\Omega/f^2=1$ leads to separability of the PT solutions 
to any order, generalizing what has been noted before in the 
case of second order PT by Martel \& Freudling (1991).
As mentioned above, the approximation 
$f(\Omega,\Lambda) \approx \Omega^{1/2}$ is actually very good 
in practice; for example, the exact solution for the 
$\Lambda=0$ case gives $D_2/(D_1)^2 =1+3/17 (\Omega^{-2/63}-1)$,
extremely insensitive to $\Omega$, even more than what the
approximation  $f(\Omega,\Lambda) = \Omega^{3/5}\approx \Omega^{1/2}$
would suggest, since for most of the time evolution $\Omega$ and
$\Lambda$ are close to their Einstein-de Sitter values.

\clearpage

\begin{figure}[t!]
\centering
\centerline{\epsfxsize=18. truecm \epsfysize=18. truecm
\epsfbox{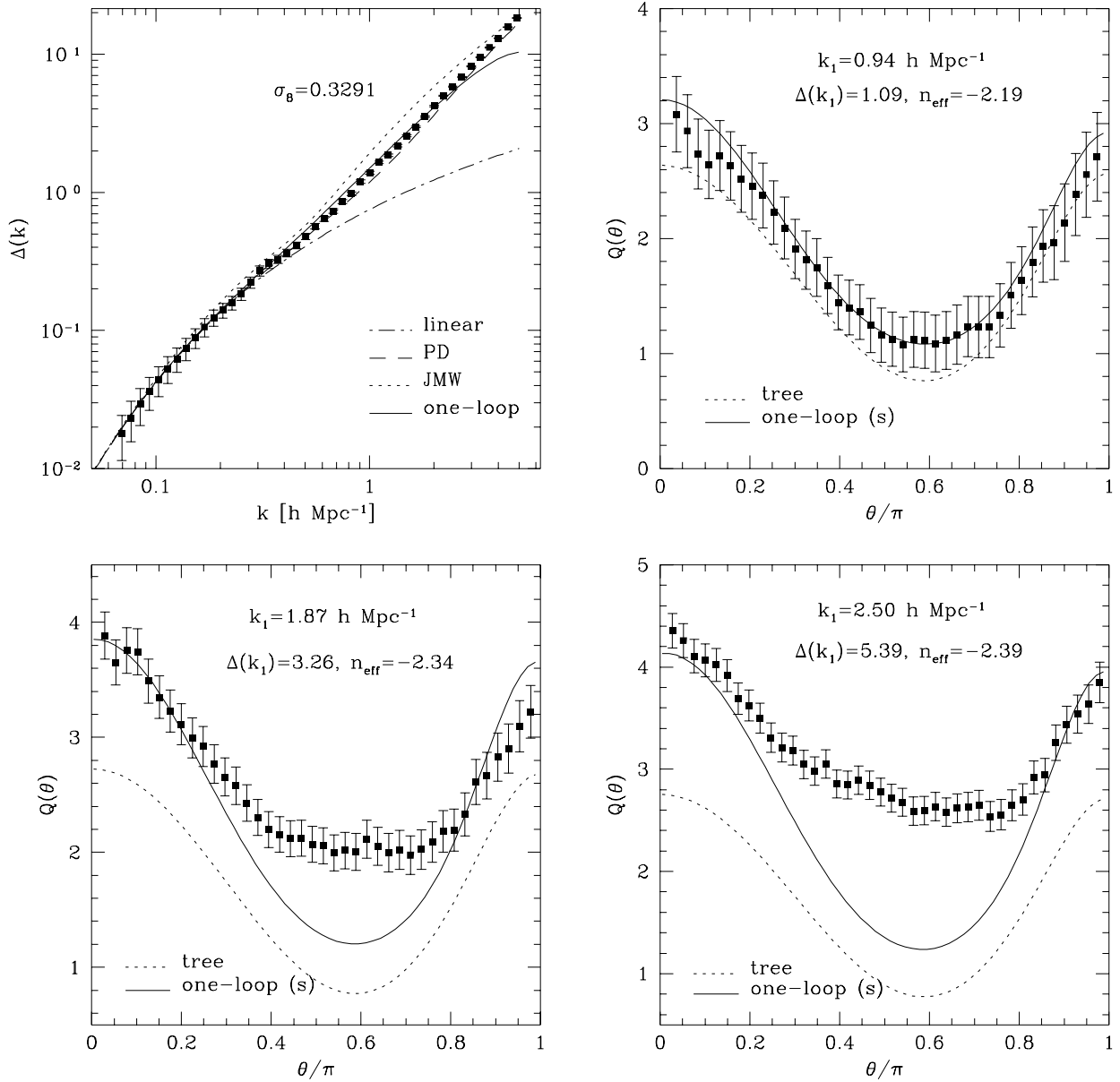}}
\caption{Same as Figure~\protect{\ref{fig6}} for a later
(more non-linear) output, $\sigma_8=0.3291$.
}
\label{fig7}
\end{figure}

\begin{figure}[t!]
\centering
\centerline{\epsfxsize=18. truecm \epsfysize=18. truecm
\epsfbox{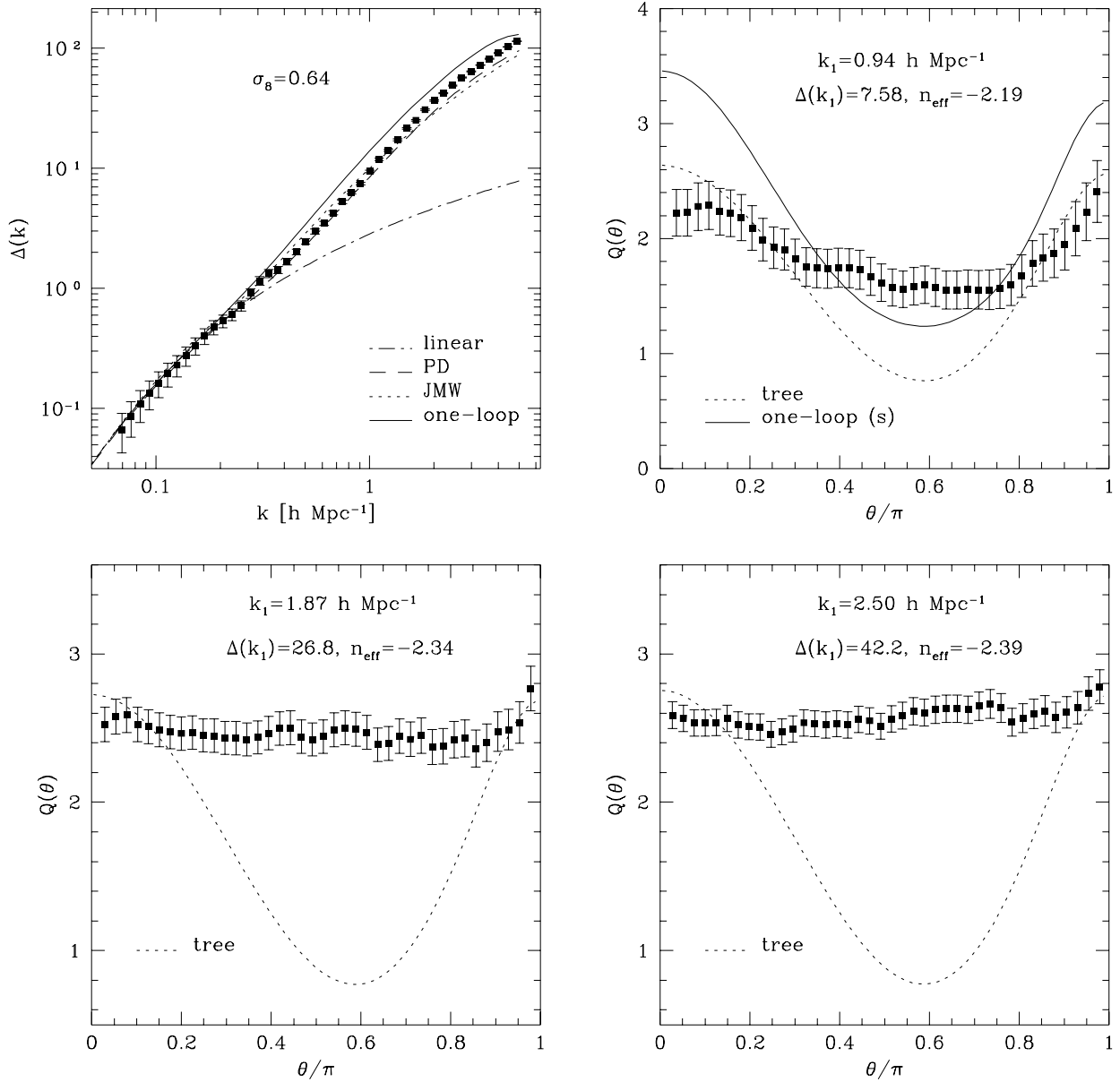}}
\caption{Same as Figure~\protect{\ref{fig6}} for the latest
(most non-linear) output, $\sigma_8=0.64$.
}
\label{fig8}
\end{figure}

\begin{figure}[t!]
\centering
\centerline{\epsfxsize=18. truecm \epsfysize=18. truecm
\epsfbox{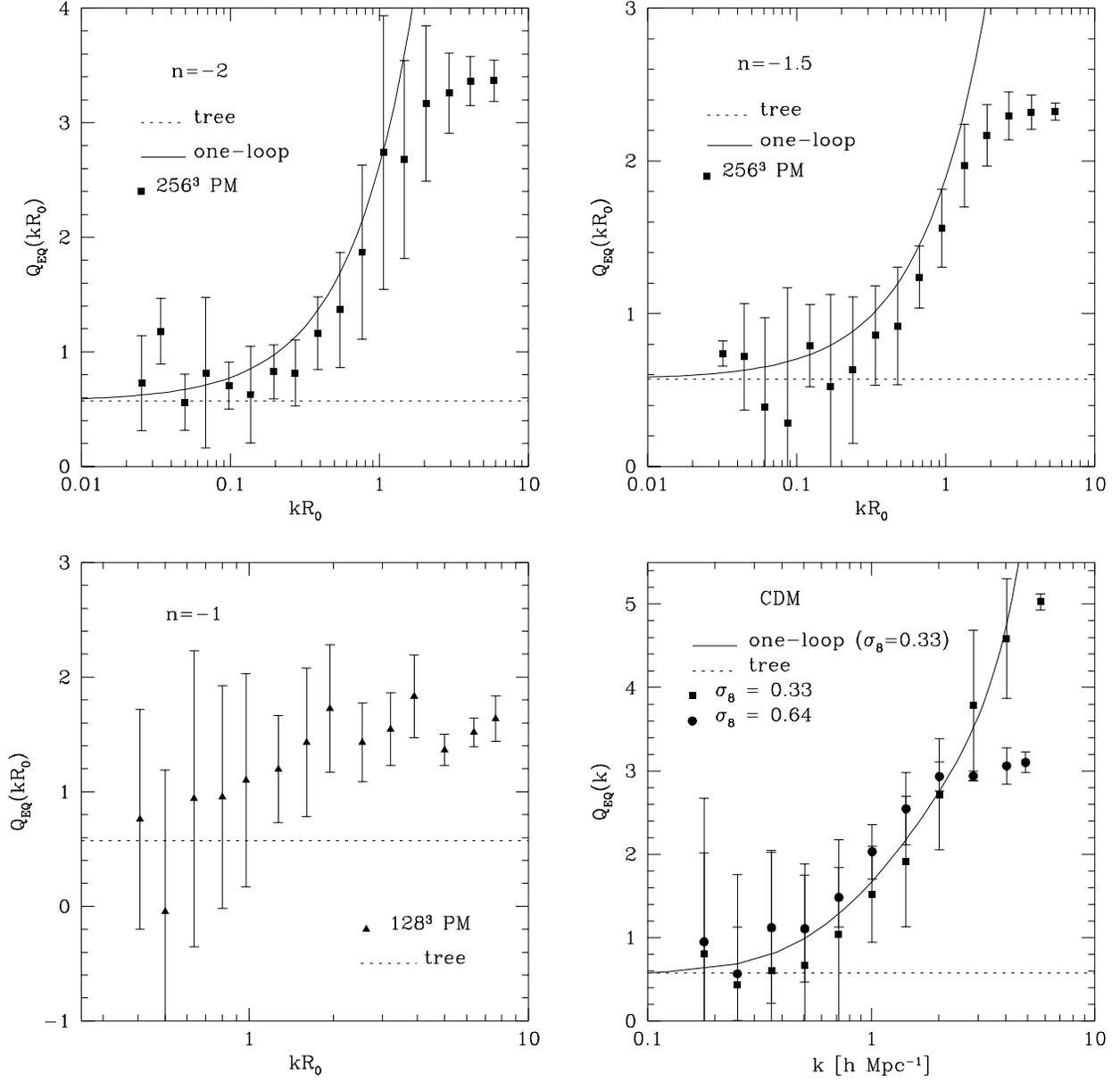}}
\caption{The hierarchical amplitude $Q$ for equilateral configurations as
a function of scale for $n=-2,-1.5,-1$ and the CDM
$\sigma_8=0.33,0.64$ outputs. 
}
\label{fig9}
\end{figure}

\begin{figure}[t!]
\centering
\centerline{\epsfxsize=13. truecm \epsfysize=13. truecm
\epsfbox{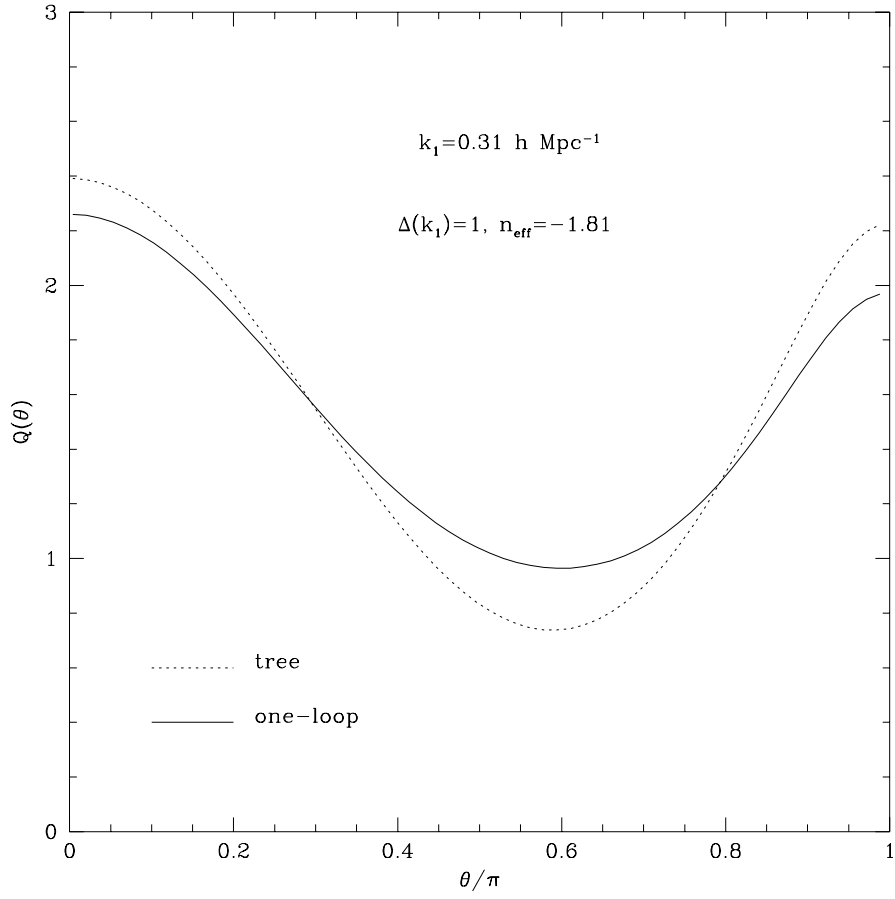}}
\caption{The hierarchical amplitude $Q$ for triangle configurations with
$k_{1}/k_{2}=2$ for  the CDM $\sigma_8=0.64$ output at
scales where $\Delta \approx 1$.
}
\label{fig10}
\end{figure}

\end{document}